\documentclass[journal,comsoc]{IEEEtran}
\IEEEoverridecommandlockouts
\usepackage[T1]{fontenc}
\usepackage{marvosym}
\usepackage{authblk}

\usepackage{amsmath}
\usepackage{comment} %
\usepackage{cite}
\usepackage{graphicx}
\usepackage{epsfig,graphics,subfigure,psfrag,amsmath,amssymb}
\usepackage{booktabs}
\usepackage{amsfonts}
\usepackage{epstopdf}
\usepackage{amssymb}
\usepackage{array}
\usepackage{amsmath}
\usepackage{makecell}
\usepackage{multirow}

\begin{document}
\title{Intelligent Angle Map-based Beam Alignment for RIS-aided mmWave Communication Networks}

\author{Hao~Xia,
        Qing~Xue,~\IEEEmembership{Member,~IEEE,}
        Yanping~Liu,
        Binggui~Zhou,
        Meng~Hua, \rm{and}
        Qianbin~Chen,~\IEEEmembership{Senior Member,~IEEE}
\thanks{Manuscript received XXX XX, 2024; revised XXX XX, 202X. This work was supported by the National Natural Science Foundation of China under Grant U23A20279.}

\thanks{Hao Xia, Qing Xue, and Qianbin Chen are with the School of Communications and Information Engineering, Chongqing University of Posts and Telecommunications, Chongqing 400065, China (e-mails: s220101167@stu.cqupt.edu.cn; xueq@cqupt.edu.cn; chenqb@cqupt.edu.cn).}

\thanks{Yanping Liu is with the School of Big Data Statistics, Guizhou University of Finance and Economics, Guiyang 550025, China (e-mail: liuyanping@mail.gufe.edu.cn).}

\thanks{Binggui Zhou and Meng Hua are with the Department of Electrical and Electronic Engineering, Imperial College London, SW7 2AZ London, U.K. (e-mails: binggui.zhou@imperial.ac.uk; m.hua@imperial.ac.uk).}
}

\maketitle

\begin{abstract}
Recently, reconfigurable intelligent surface (RIS) has been widely used to enhance the performance of millimeter wave (mmWave) communication systems, making beam alignment more challenging.
To ensure efficient communication, this paper proposes a novel intelligent angle map-based beam alignment scheme for both general user equipments (UEs) and RIS-aided UEs simultaneously in a fast and effective way. Specifically, we construct a beam alignment architecture that utilizes only angular information. To obtain the angle information, the currently hottest seq2seq model -- the Transformer -- is introduced to offline learn the relationship between UE geographic location and the corresponding optimal beam direction. Based on the powerful machine learning model, the location-angle mapping function, i.e., the angle map, can be built. As long as the location information of UEs is available, the angle map can make the acquisition of beam alignment angles effortless. In the simulation, we utilize a ray-tracing-based dataset to verify the performance of the proposed scheme. It is demonstrated that the proposed scheme can achieve high-precision beam alignment and remarkable system performance without any beam scanning.
\end{abstract}

\begin{IEEEkeywords}
Millimeter wave communications, reconfigurable intelligent surface, beam alignment, deep learning.
\end{IEEEkeywords}

\IEEEpeerreviewmaketitle

\setlength{\parskip}{0pt}

\section{Introduction}
As one of the core technologies of 5G, millimeter wave (mmWave) communication plays a key role in delivering high data rates and low latency. The frequency range of mmWave is approximately 30 GHz to 300 GHz, offering abundant spectrum resources that can support large-scale data transmission. However, there are many technical challenges in effectively applying mmWave to practical systems. The short wavelength of mmWave results in weak penetration through obstacles, limiting the signal propagation distance and making mmWave communication systems difficult to achieve wide-area signal coverage. For the same reason, signals transmitted through mmWave can be easily blocked by obstacles, which greatly affects the reliability of communication. To compensate for the disadvantages mentioned above, beamforming technology is typically utilized in mmWave communication networks. Beamforming can help adjusting the phase and amplitude of transmitted signals from multiple antenna elements to concentrate the energy in a specific direction, thereby enhancing signal strength and reducing interference. In recent years, reconfigurable intelligent surface (RIS) is an emerging technology that can intelligently control the phase, amplitude, and other characteristics of passive reflecting elements on the surface to manipulate propagation paths of wireless signals. Thus, RIS is usually introduced into mmWave communication systems to optimize signal propagation characteristics and expand signal coverage. While RIS brings effective gains, it also poses challenges for beam alignment. This is because the RIS-assisted system requires joint management of two-hop transmissions, whereas traditional systems only need to consider single-hop transmission topologies. Beam alignment is capable of ensuring that signal beams from the transmitter and receiver are aligned precisely in physical space, which plays a critical role in enhancing system performance in highly directional communication systems \cite{10422712}. Especially, beam alignment can help maximizing signal strength while minimizing errors and delays, thus not only improving the signal quality for user equipment (UE) but also effectively reducing the intra-cell interference.
\par In this paper, we present a location-aware deep learning-enabled beam alignment scheme. To be specific, a novel Transformer-based angle map (AM) and a novel AM-based beam alignment method for RIS-assisted mmWave communication system are proposed, aiming at reducing beam alignment overhead and facilitating multi-UEs beam alignment in RIS-aided networks. Due to the UE blockage detection and the phase shift design of the RIS, the conventional optimization-based beam alignment schemes will suffer extra inevitable complexity when handling these issues. With the number of UEs increases, there may be even complexity inflation caused by massive optimization parameters. However, the proposed AM-based beam alignment scheme will not trapped by such problem and even suffice for achieving fast and accurate beam alignment for multi-UEs (including both general UEs and RIS-aided UEs). Particularly, the AM is a kind of mapping function between UEs' location and beam alignment angles within a specific area. Utilizing the AM to forecast optimal beam directions can greatly cut the time and overhead brought by beam sweeping compared to search-based beam alignment schemes.

\subsection{Related Works}
The exhaustive search algorithm is the most basic and straightforward beam alignment method \cite{8458146}. To find the optimal beam direction, the exhaustive search algorithm needs to systematically traverse all possible beam directions and evaluate the performance of each direction based on signal quality indicator (e.g. received signal strength, channel quality indicator, etc.). To implement the exhaustive search, firstly, it is necessary to define a global codebook that includes all spatial angles based on the configuration of the antenna array on the transmitter. Then, the mmWave base station (mBS) and the UE adjust their antenna arrays sequentially, attempting each predefined beam direction in the codebook and recording signal quality for each direction set. According to the collected signal quality indicator, the best combination of directions will be selected as the practical transmission/reception direction for the mBS and the UE. In general, the exhaustive search is relatively simple, requiring no complex algorithms or models and no prior information. It can achieve high accuracy in most static scenes and be smoothly applied to RIS-assisted communication scenarios \cite{10422712}. However, since the exhaustive search requires scanning all possible beam directions one by one, the exorbitant time and energy cost are apparent, which makes the algorithm non-cost-efficient in dynamic environments or multi-UE scenarios. To alleviate such overhead, hierarchical beam search is introduced subsequently \cite{7947209,7390101,9129778,9325920}. In particular, hierarchical search typically begins with wide beams to perform a coarse search across entire space with a large step size to identify a rough region. Then, within the found region, narrow beams are used for more detailed scanning. By repeating the procedure, the search range will gradually narrow down until the optimal beam direction is found. As long as the number of searching layers and the beam width for each layer are properly configured, hierarchical search can achieve effective beam alignment with significantly reduced search time and computation. Hierarchical search is also widely used to achieve beam alignment for RIS-assisted communication systems \cite{9129778,9325920}. In \cite{9129778}, a multi-beam training method is proposed, which divides the RIS into several sub-arrays and simultaneously designs multi-beam steering in different directions. Users then select optimal beams based on the received signal strength. A hierarchical codebook-based cooperative beam training scheme is proposed to perform the cascade channel estimation in \cite{9325920}. Besides the mentioned scanning-based beam alignment algorithms, traditional beam alignment algorithms also include the multiple signal classification and the estimation of signal parameters via rotational invariance techniques algorithm, both of which achieve beam alignment through estimating the angle of arrival (AoA) and angle of departure (AoD) of signals based on geometric models \cite{7938435,8846224}. There are also compressed sensing algorithms \cite{8114345,9573459}, which exploits the sparsity of channel for beam alignment, as well as strategies that dynamically adjust beam directions based on feedback information, such as channel quality indicators and channel state information (CSI) \cite{6884811,10038557}.
\par Although traditional beam alignment schemes generally have low complexity and can be easily implemented, most of them are still inherently unsuitable for dynamic environments since they mainly rely on fixed algorithms and heuristics. In order to overcome such limitation, machine learning (ML)-based solutions are explored \cite{8395053,10542373}. ML models can extract useful features and patterns from vast amounts of data, enabling intelligent decision-making. Therefore, the ML-based beam alignment is able to achieve dynamic environment adaptation and complex feature processing, making the beam alignment process more efficient, accurate, and flexible \cite{9690703,10241295,10151679,10130629,10505148}. An end-to-end neural network framework is proposed to jointly learn the probing codebook and the beam predictor in \cite{9690703}. \cite{10241295} , \cite{10151679} and \cite{10130629} utilize deep learning to facilitate beam alignment. In \cite{10241295}, hierarchical beam alignment schemes utilizing deep learning-based probing codebooks are proposed for multiple-input single-output (MISO) and multiple-input multiple-output (MIMO) systems respectively. A grid-free beam alignment method that needs no quantized codebooks is proposed in \cite{10151679}, which achieves beam alignment through deep learning-based site-specific probing beams. In \cite{10130629}, the extremely large-scale RIS is considered in the near-field communication, and two deep learning-based beam alignment schemes are proposed to reduce the overhead and the number of codewords in the codebook. An indoor RIS-aided communication scene is studied in \cite{10505148}, where beam alignment is effectively achieved through the proposed codebook-based 3D beam scanning scheme.
\par Most of beam alignment methods mentioned above either use beam scanning, the CSI, or feedback information to achieve beam alignment, while a small fraction pay attention to environmental information. Since the trend of the integration of sensing and communication is foreseeable, utilizing sensing-related information or sensing devices for beam alignment can better adapt to future communication environments and provide higher efficiency. Beam alignment based on environmental information typically uses external sensing data, such as location awareness, radar or light detection and ranging data, camera images, or even historical communication records between the BS and UE, to find the optimal beam direction. Both \cite{10438390} and \cite{9209195} employ radar sensing technology and use the sensing information to assist communication systems in achieving beam alignment. Location information is leveraged in \cite{9832551} and \cite{10574385} for achieving beam alignment. In \cite{9832551}, a location and orientation information-enabled deep neural network-based beam alignment scheme is proposed. Channel knowledge map, which can obtain the required information for beam alignment based on the location information of the receiver, is proposed to conduct beam alignment in \cite{10574385}. Two computer vision-based beam alignment methods are developed in \cite{10218358} and \cite{10100676}. In \cite{10218358}, the prediction of candidate beam sets is carried out through the visual information. \cite{10100676} achieves the beam tracking of RIS through identifying the location of the transmitter and the direction of the receiver, which are both extracted from the visual information.

\subsection{Main Contributions}
To efficiently address the multi-UE beam alignment challenge in RIS-assisted mmWave communications, this paper proposes a novel approach that introduces an intelligent AM. The proposed AM is generated based on the Transformer model, which can effectively learn the mapping features between the location and angle information, and is adept in handling parallel input data. Such characteristics enable the proposed AM to be capable of performing fast and high-precision optimal beam direction prediction for multi-UEs. It is worth saying that the proposed beam alignment scheme achieves multi-UEs alignment fully relying on the angle information. Thus, the extra complexity introduced by RIS will have little impact on the proposed scheme. Several key features of the proposed method are listed as follows.
\begin{itemize}
  \item \textbf{Transformer-based angle estimation architecture.} A Transformer-based intelligent AM is built to forecast corresponding beam alignment angles based on the location information of UEs. The Transformer model has the advantage of processing input data in parallel, allowing the intelligent AM to simultaneously predict optimal beam directions for multiple line-of-sight (LoS) and non-LoS (NLoS) UEs.
  \item \textbf{Multi-UEs joint beam alignment method.} A multi-UEs joint beam alignment method for RIS-auxiliary mmWave system is proposed. By introducing a series of transformation operations, the proposed method can achieve simultaneous beam alignment for both LoS UEs and NLoS UEs merely through the angle information without any beam scanning or traditional beamforming optimizing. Therefore, enabled by the proposed AM, the proposed beam alignment scheme can achieve fast and accurate multi-UEs beam alignment in a low-cost way.
  \item \textbf{Capable applying to 5G/6G scenario.} The only prior information required by the proposed algorithm is the UEs' location information, which is typically obtained through localization. In the 6G vision of integrated sensing and communication networks, sensing information will be easily shared by communication modules and utilized in communication functions. Therefore, the proposed scheme is considered to be applied to 6G. In fact, as long as the location information is attainable and the amount of data is sufficient, the proposed scheme can be applied to a wide variety of communication scenarios to achieve effective beam alignment.
  \item \textbf{Fast and effective beam alignment with reduced overhead.} Since the proposed scheme is implemented in a data-driven way, it does not require any beam scanning operations for beam alignment. Moreover, for most of the traditional beam alignment algorithms, the overall overhead of the system and complexity will increase sharply with the number of UEs in RIS-assisted multi-UEs scenarios. In contrast, the proposed algorithm is inherently good at handling multi-UEs problem and can simultaneously align both direct-link UEs and RIS-assisted UEs. As a result, significant computational and time overhead reduction can be achieved by the proposed alignment method.
  \item \textbf{Numerical simulation.} Based on the widely used public dataset DeepMIMO, which is obtained using the ray-tracing software Wireless InSite \cite{remcom2019}, we train the proposed AM and verify the system performance of the proposed algorithm. The comparison with other advanced algorithms is also implemented. Simulation results show the proposed scheme can achieve impressive performance with a well-trained AM.
\end{itemize}

\subsection{Paper Organization}
The rest of this work is organized as follows. The system model and the proposed angle-based beam alignment scheme are described in Section II. The Transformer-based angle map is explained in Section III. The numerical simulations are conducted in Section IV. Finally, the conclusion is provided in Section V.

\section{System Model}
\begin{figure}[t!]
	\centering
	\includegraphics[width=0.45\textwidth]{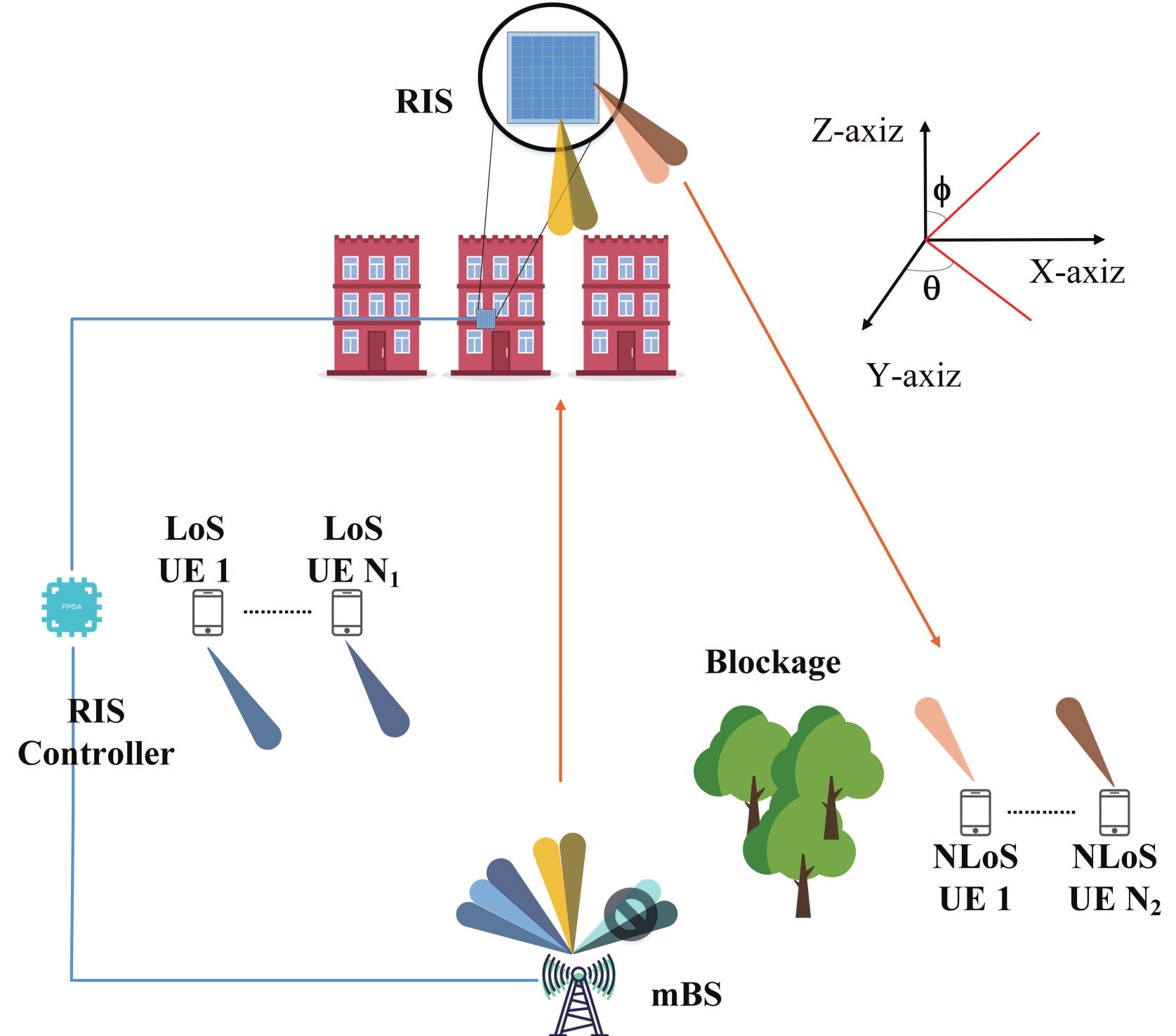}
	\caption{RIS-assisted mmWave communication system.}
	\label{fig:1}
\end{figure}
Fig. 1 shows a RIS-aided mmWave MIMO system, where a single mmWave base station serves $N$ UEs (which consist of $N_1$ unblocked UEs and $N_2$ blocked UEs) and the RIS is deployed on the wall to establish mBS-RIS-UE links. The mBS is equipped with a uniform planar array (UPA) of ${N_T} = N_x^T \times N_z^T$, where $x$ and $z$ represent the $x$-axis and $z$-axis respectively. Each UE is equipped with the same UPA of ${N_R} = N_x^R \times N_z^R$. The RIS consists of $M = {M_x} \times {M_z}$ passive reflection elements. Table I provides the main parameter definitions.
\begin{table}
\renewcommand{\arraystretch}{1.2}
    \begin{center}
    \caption{Notation of Important Variables}
    \begin{footnotesize}
    \begin{tabular}{c|c} 
      \hline
      \textbf{Symbol} & \textbf{Definition} \\
      \hline
      $N_T$ & Number of antennas at mBS \\
      \hline
      $N_R$ & Number of antennas at UE \\
      \hline
      $M$ & Number of RIS reflection elements \\
      \hline
      $N$ & Number of UEs \\
      \hline
      $N_1$ & Number of unblocked UEs \\
      \hline
      $N_2$ & Number of blocked UEs \\
      \hline
      ${{\vartheta _n}}$ & Blocking coefficient of UE $n$ \\
      \hline
      ${\textbf{H}_{n,\textrm{LoS}}}$ & LoS channel between mBS and UE $n$ \\
      \hline
      $\textbf{H}_{n,\textrm{NLoS}}$ & NLoS channel between mBS and UE $n$ \\
      \hline
      ${\textbf{h}_{\textrm{NLoS},1}}$ & LoS channel between mBS and RIS \\
      \hline
      ${\textbf{h}_{n,\textrm{NLoS},2}}$ & LoS channel between RIS and UE $n$ \\
      \hline
      ${\bar {\textbf{h}}_{\textrm{NLoS},1}}$ & LoS component of ${\textbf{h}_{\textrm{NLoS},1}}$ \\
      \hline
      ${\bar {\textbf{h}}_{n,\textrm{NLoS},2}}$ & LoS component of ${\textbf{h}_{n,\textrm{NLoS},2}}$ \\
      \hline
      ${{\theta _{R,l}}}({{\varphi _{R,l}}})$ & Azimuth (Pitch) AoA at UE of $l$-th path \\
       \hline
      ${{\theta _{T,l}}}({{\varphi _{T,l}}})$ & Azimuth (Pitch) AoD from mBS of $l$-th path \\
      \hline
      ${{\theta _{in,l}}}({{\varphi _{in,l}}})$ & Azimuth (Pitch) AoA at RIS of $l$-th path \\
      \hline
      ${{\theta _{out,l}}}({{\varphi _{out,l}}})$ & Azimuth (Pitch) AoD from RIS of $l$-th path \\
      \hline
    \end{tabular}
    \end{footnotesize}
    \end{center}
\end{table}
\subsection{Channel Model}
The transmit signal from the mBS to UE $n\left( {n \in {\cal N},{\cal N} = \left\{ {1,2,...,N} \right\}} \right)$ is defined as ${s_n} \in \mathbb{C}$, which satisfies the average power constraint $\mathbb{E}\left[ {\left| {{s_n^2}} \right|} \right] = 1$. Since there are two UE types in the scenario (i.e., unobstructed UE and obstructed UE), the receive signal of UE $n$ is written as
\begin{equation}
\begin{split}
& {\textbf{y}_n} = \sqrt {{P_n}} \textbf{w}_n^H\left[ {{\vartheta _n}\textbf{H}_{n,LoS}^H + \left( {1 - {\vartheta _n}} \right)\textbf{H}_{n,NLoS}^H} \right]{\textbf{f}_n}{s_n} \\
& + \sum\limits_{m \ne n}^N {\sqrt {{P_m}} \textbf{w}_n^H\left[ {{\vartheta _n}\textbf{H}_{n,LoS}^H + \left( {1 - {\vartheta _n}} \right)\textbf{H}_{n,NLoS}^H} \right]{\textbf{f}_m}{s_m}} \\ & + \textbf{w}_n^H\eta,
\end{split}
\end{equation}
where ${\vartheta _n} \in \left\{ {0,1} \right\}$ denotes the channel blocking coefficient, ${\vartheta_n}  = 1$ means that there is a LoS link between the mBS and UE $n$, ${\vartheta_n}  = 0$ implies that only NLoS links exist. ${P_n}$ is the transmit power for UE $n$, $\eta \sim {{\cal N}_\mathbb{C}}\left( {0,{\sigma ^2}} \right)$ represents the Gaussian white noise. ${\textbf{f}_n} \in {\mathbb{C}^{{N_T} \times 1}}$ denotes the transmit beamforming vector and ${\textbf{w}_n} \in {\mathbb{C}^{{N_R} \times 1}}$ denotes the receive beamforming vector. The LoS channel between the mBS and UE $n$ is
\begin{equation}
{{\bf{H}}_{n,{\rm{LoS}}}} = \sum\limits_{l = 1}^L {{\alpha _l}{{\bf{a}}_R}\left( {{\theta _{R,l}},{\varphi _{R,l}}} \right){\bf{a}}_T^H\left( {{\theta _{T,l}},{\varphi _{T,l}}} \right)} ,
\end{equation}
where $L$ denotes the number of signal paths between the mBS and UE $n$, $l = 1$ denotes the LoS path, ${\alpha _l}$ is the channel complex gain. ${\theta _{R,l}}$ and ${\varphi _{R,l}}$ are the azimuth and pitch AoA at UE respectively, ${\theta _{T,l}}$ and ${\varphi _{T,l}}$ are the azimuth and pitch AoD at mBS respectively. ${{\bf{a}}_R}\left(  \cdot  \right)$ and ${{\bf{a}}_T}\left(  \cdot  \right)$ denote the receive and transmit array steering vectors. If UE $n$ is NLoS UE, the channel between the mBS and UE $n$ will become a cascade channel, which can be defined as
\begin{equation}
  {\bf{H}}_{n,{\rm{NLoS}}}^{} = {{\bf{h}}_{n,{\rm{NLoS}},2}}{{\bf{\Phi }}}{{\bf{h}}_{{\rm{NLoS}},1}},
\end{equation}
where ${{\bf{\Phi }}} \in {\mathbb{C}^{M \times M}}$ denotes the RIS reflection matrix, ${{\bf{h}}_{{\rm{NLoS}},1}} \in {\mathbb{C}^{M \times {N^{T}}}}$ and ${{\bf{h}}_{n,{\rm{NLoS}},2}} \in {\mathbb{C}^{{N^{R}} \times M}}$ are LoS channel matrices between the mBS and the RIS, and between the RIS and UE $n$ respectively, defined as
\begin{equation}
  {{\bf{h}}_{{\rm{NLoS}},1}} = \textstyle\sum_{l = 1}^{{L}} {{\alpha _l}{{\bf{a}}_R}\left( {{\theta _{in,l}},{\varphi _{in,l}}} \right){\bf{a}}_T^H\left( {{\theta _{T,l}},{\varphi _{T,l}}} \right)} ,
\end{equation}
\begin{equation}
  {{\bf{h}}_{n,{\rm{NLoS}},2}} = \textstyle\sum_{l = 1}^{{L}} {{\alpha _l}{{\bf{a}}_R}\left( {{\theta _{R,l}},{\varphi _{R,l}}} \right){\bf{a}}_T^H\left( {{\theta _{out,l}},{\varphi _{out,l}}} \right)} ,
\end{equation}
where ${\theta _{in,l}}$ and ${\varphi _{in,l}}$ are the azimuth and pitch AoA, ${\theta _{out,l}}$ and ${\varphi _{out,l}}$ are the azimuth and pitch AoD at RIS, respectively. In addition, the array steering vectors involved above are denoted as
\begin{equation}
{{\bf{a}}_R}\left( {{\theta _{R,l}},{\varphi _{R,l}}} \right) = {\bf{a}}\left( {{\theta _{R,l}},{\varphi _{R,l}},{N^R}} \right),
\end{equation}
\begin{equation}
{{\bf{a}}_T}\left( {{\theta _{T,l}},{\varphi _{T,l}}} \right) = {\bf{a}}\left( {{\theta _{T,l}},{\varphi _{T,l}},{N^T}} \right),
\end{equation}
\begin{equation}
{{\bf{a}}_R}\left( {{\theta _{in,l}},{\varphi _{in,l}}} \right) = {\bf{a}}\left( {{\theta _{in,l}},{\varphi _{in,l}},M} \right),
\end{equation}
\begin{equation}
{{\bf{a}}_T}\left( {{\theta _{out,l}},{\varphi _{out,l}}} \right) = {\bf{a}}\left( {{\theta _{out,l}},{\varphi _{out,l}},M} \right),
\end{equation}
where $\textbf{a}\left( \cdot \right)$ is the general array steering vector written as
\begin{equation}
\textbf{a}\left( {\theta ,\varphi ,{N^{ant}}} \right) = {\textbf{a}_x}\left( {\theta ,\varphi ,{N_x^{ant}}} \right) \otimes {\textbf{a}_z}\left( {\varphi ,{N_z^{ant}}} \right),
\end{equation}
where ${N^{ant}} = {N_x^{ant}} \times {N_z^{ant}}$ denotes the number of antennas on the UPA. The concrete expressions of array steering vectors on the $x$-axis and the $z$-axis are ${\textbf{a}_x}\left( {\theta ,\varphi ,{N_x^{ant}}} \right) = {\left[ {1,{e^{jksin\left( \theta  \right)\sin \left( \varphi  \right)}},...,{e^{jk\left( {{N_x^{ant}} - 1} \right)sin\left( \theta  \right)\sin \left( \varphi  \right)}}} \right]^T}$ and ${\textbf{a}_z}\left( {\varphi ,{N_z^{ant}}} \right) = {\left[ {1,{e^{jk\cos \left( \varphi  \right)}},...,{e^{jk\left( {{N_z^{ant}} - 1} \right)\cos \left( \varphi  \right)}}} \right]^T}$, where $k = \frac{{2\pi d}}{\lambda }$, $\lambda$ is the wavelength and $d$ is the antenna spacing. In (3), the RIS reflection matrix ${\bf{\Phi }} = diag\left( {{e^{j{\phi _{1,1}}}},...,{e^{j{\phi _{{M_x},{M_z}}}}}} \right)$, where ${\phi _{{m_x},{m_z}}} \in \left[ {0,2\pi } \right]$ is the phase shift of the ${m_x}$-th row ${m_z}$-th column RIS reflection element $\left( {{m_x} \in \left\{ {1,...,{M_x}} \right\},{m_z} \in \left\{ {1,...,{M_z}} \right\}} \right)$.

In addition, commonly used traditional beamforming algorithms all correlate with channel matrix \cite{7400949}, and channel can be seen as a mapping of angles and RIS phase as the above formula derivation shows. Hence, the transmit and receive beamforming vectors in (1) can be also regarded as functions of angles and RIS phases:
\begin{equation}
\begin{split}
& \textbf{f} = \textbf{f}\left( \textbf{H} \right),\textbf{w} = \textbf{w}\left( \textbf{H} \right),\textbf{H} = \textbf{H}\left( {\theta ,\varphi ,\mathbf{\Phi} } \right) \\
& \Rightarrow \textbf{f} = \textbf{f}\left( {\theta ,\varphi ,\mathbf{\Phi} } \right), \textbf{w} = \textbf{w}\left( {\theta ,\varphi ,\mathbf{\Phi} } \right).
\end{split}
\end{equation}
\begin{small}
\begin{equation}
\begin{split}
& SIN{R_n}\\
& = \frac{{{{\left| {{\vartheta _n}\textbf{w}_n^H\textbf{H}_{n,\textrm{LoS}}^H{\textbf{f}_n} + \left( {1 - {\vartheta _n}} \right)\textbf{w}_n^H\textbf{H}_{n,\textrm{NLoS}}^H{\textbf{f}_n}} \right|}^2}{P_n}}}{{\sum\limits_{m \ne n}^N {{{\left| {{\vartheta _n}\textbf{w}_n^H\textbf{H}_{n,\textrm{LoS}}^H{\textbf{f}_m} + \left( {1 - {\vartheta _n}} \right)\textbf{w}_n^H\textbf{H}_{n,\textrm{NLoS}}^H{\textbf{f}_m}} \right|}^2}{P_m}}  + \sigma _w^2}} \\
& = \frac{{{{\left| {\textbf{w}_n^H{\mathbf{\Omega} _n}{\textbf{f}_n}} \right|}^2}{P_n}}}{{\sum\limits_{m \ne n}^N {{{\left| {\textbf{w}_n^H{\mathbf{\Omega} _n}{\textbf{f}_m}} \right|}^2}{P_m}}  + \sigma _w^2}}.
\end{split}
\end{equation}
\end{small}
\par The signal-interference-noise ratio (SINR) of UE $n$ is denoted in (12), where ${\mathbf{\Omega} _n} = {\vartheta _n}\textbf{H}_{n,\textrm{LoS}}^H + \left( {1 - {\vartheta _n}} \right)\textbf{H}_{n,\textrm{NLoS}}^H$ and $\sigma _w^2$ is the variance of $\textbf{w}_n^H\eta$. Moreover, the achievable rate of UE $n$ is ${R_n} = \log_{2} \left( {1 + SIN{R_n}} \right)$.
\subsection{Problem Formulation and Transformation}
The system sum rate can be described as
\begin{equation}
  R = \sum\limits_{n = 1}^N {\log_2 \left( {1 + \frac{{{{\left| {{\bf{w}}_n^H{{\bf{\Omega }}_n}{{\bf{f}}_n}} \right|}^2}{P_n}}}{{\sum\limits_{m \ne n}^N {{{\left| {{\bf{w}}_n^H{{\bf{\Omega }}_n}{{\bf{f}}_m}} \right|}^2}{P_m}}  + \sigma _w^2}}} \right)} .
\end{equation}
\par The objective of this paper is to maximize the sum rate through beam alignment. The corresponding optimization problem is
\begin{equation}
\begin{array}{ll}
& \mathop {\max }\limits_{{\bf{\Theta }},{\bf{\Psi }},{{\bf{\Theta }}_{RIS}},{{\bf{\Psi }}_{RIS}},{\bf{\Phi }},{\bf{P}}} R, \\
& {\rm{s.t.}}\left\{ \begin{array}{l}
{\rm{C1}}:{{\vartheta _n} \in \left\{ {0,1} \right\},\forall n \in {\cal N},}\\
{\rm{C2}}:{0 \le {\theta _{R,l}},{\theta _{T,l}},{\theta _{in,l}},{\theta _{out,l}} \le 2\pi ,\forall l \in {\cal L} ,}\\
{\rm{C3}}:{0 \le {\varphi _{R,l}},{\varphi _{T,l}},{\varphi _{in,l}},{\varphi _{out,l}} \le 2\pi ,\forall l \in {\cal L} ,}\\
{\rm{C4}}:{0 \le {\phi _{{m_x},{m_z}}} \le 2\pi ,\forall {m_x} \in {{\cal M}_x},{m_z} \in {{\cal M}_z},}\\
{\rm{C5}}:{0 < \sum\limits_{n = 1}^N {{P_n}}  \le {P^{\max }},}\\
{\rm{C6}}:{\left| {{{\bf{f}}_n}} \right|^2} = 1 ,\forall n \in {\cal N}, \\
{\rm{C7}}:{\left| {{\textbf{w}_n}} \right|^2} = 1 ,\forall n \in {\cal N},
\end{array} \right.
\end{array}
\end{equation}
where $\textbf{P} = \left\{ {{P_1},...,{P_N}} \right\}$ denotes the transmit power set of all UEs, ${P^{\max }}$ denotes the maximum transmit power of the mBS. ${\cal L}  = \left\{ {1,...,L } \right\}$ denotes the set of transmit paths. ${{\cal M}_x} = \left\{ {1,...,{M_x}} \right\},{{\cal M}_z} = \left\{ {1,...,{M_z}} \right\}$ are sets of the number of RIS reflection elements at each row and column. ${\bf{\Theta }} = \left\{ {{{\boldsymbol{\theta}}_1},...,{{\boldsymbol{\theta}}_N}} \right\} \in {\mathbb{C}^{N \times L}}$ and ${{\bf{\Theta }}_{RIS}} = \left\{ {{{\boldsymbol{\theta }}_1^{RIS}},...,{{\boldsymbol{\theta }}_{N_2}^{RIS}}} \right\} \in {\mathbb{C}^{N_2 \times L}}$ are respectively the azimuth angle matrix for all UEs and the azimuth angle matrix related to RIS-assisted UEs, ${\bf{\Psi }} = \left\{ {{{\boldsymbol{\varphi }}_1},...,{{\boldsymbol{\varphi }}_N}} \right\} \in {\mathbb{C}^{N \times L}}$ and ${{\bf{\Psi }}_{RIS}} = \left\{ {{{\boldsymbol{\varphi }}_1^{RIS}},...,{{\boldsymbol{\varphi }}_{N_2}^{RIS}}} \right\} \in {\mathbb{C}^{N_2 \times L}}$ are pitch angle matrices with similar denotation to ${\bf{\Theta }}$ and ${{\bf{\Theta }}_{RIS}}$, ${{\boldsymbol{\theta }}_n} = \left( {{\theta _{R,1}},{\theta _{T,1}},...,{\theta _{R,L}},{\theta _{T,L}}} \right)$ and ${{\boldsymbol{\varphi }}_n} = \left( {{\varphi _{R,1}},{\varphi _{T,1}},...,{\varphi _{R,L}},{\varphi _{T,L}}} \right)$ are azimuth and pitch angle vectors of UE $n$, ${{\boldsymbol{\theta }}_v^{RIS}} = \left( {{\theta _{in,1}},{\theta _{out,1}},...,{\theta _{in,L}},{\theta _{out,L}}} \right)$ and ${{\boldsymbol{\varphi }}_v^{RIS}} = \left( {{\varphi _{in,1}},{\varphi _{out,1}},...,{\varphi _{in,L}},{\varphi _{out,L}}} \right)$ are azimuth and pitch angle vectors of NLoS UE $v, {v \in \left\{ {1,...,N_2} \right\}}$.

Because of the discrete blocking coefficient ${\vartheta _n}$, the above optimization problem is a non-convex problem which is difficult to solve. We will convert it into an easily solvable form. In fact, the phase of each reflection element on the RIS is deeply related to the angle information \cite{10172306,8746155}. Take a single NLoS UE as an example, by maximizing the spectral efficiency of the NLoS link between mBS and the blocked UE $v,v \in {\cal \mathcal{N}}$, the optimal reflection matrix is presented by
\begin{equation}
  \mathbf{\Phi} ^{{\rm{opt}}} = \mathop {\arg\max }\limits_{{\mathbf{\Phi}}} {\left\| {{{\bar {\textbf{h}}}_{v,\textrm{NLoS},2}}{\mathbf{\Phi}}{{\bar {\textbf{h}}}_{\textrm{NLoS},1}}} \right\|^2},
\end{equation}
where ${\bar {\textbf{h}}_{\textrm{NLoS},1}}$ and ${\bar {\textbf{h}}_{v,\textrm{NLoS},2}}$ denote the LoS component of ${\textbf{h}_{\textrm{NLoS},1}}$ and ${\textbf{h}_{v,\textrm{NLoS},2}}$ respectively. Substituting (4) and (5) into (15) and denoting $\chi  = \textbf{a}_M^H\left( {{\theta _{out,1}},{\varphi _{out,1}}} \right){\mathbf{\Phi}}\textbf{a}_M^{}\left( {{\theta _{in,1}},{\varphi _{in,1}}} \right)$, we can obtain
\begin{equation}
  \mathbf{\Phi}^{{\rm{opt}}} = \mathop {\arg\max }\limits_{{\mathbf{\Phi}}} {\left| \chi  \right|^2} {\left\| {\textbf{a}_{{R}}^{}\left( {{\theta _{R,1}},{\varphi _{R,1}}} \right)}{\textbf{a}_{{T}}^H\left( {{\theta _{T,1}},{\varphi _{T,1}}} \right)} \right\|^2},
\end{equation}
where ${\left\| {\textbf{a}_{{R}}^{}\left( {{\theta _{R,1}},{\varphi _{R,1}}} \right)}{\textbf{a}_{{T}}^H\left( {{\theta _{T,1}},{\varphi _{T,1}}} \right)} \right\|^2}$ is a constant. For $\textbf{a}_M^{}\left( {\theta ,\varphi } \right) = \textbf{a}\left( {\theta ,\varphi ,M} \right) = \textbf{a}\left( {\theta ,\varphi ,{M_x}} \right) \otimes \textbf{a}\left( {\theta ,\varphi ,{M_z}} \right),$ we can derive the specific expression of $\chi $ in (17) ($0 \le \left| \chi \right| \le M$).
\begin{figure*}
\begin{small}
\begin{align}
\begin{split}
  \chi = \sum\limits_{{m_x}}^{{M_x}} {\sum\limits_{{m_z}}^{{M_z}} {\exp \left\{ {jk\left[ {\left( {{m_z} - 1} \right)\left( {\cos {\varphi _{in,1}} - \cos {\varphi _{out,1}}} \right) + \left( {{m_x} - 1} \right)\left( {\sin {\theta _{in,1}}\sin {\varphi _{in,1}} - \sin {\theta _{out,1}}\sin {\varphi _{out,1}}} \right)} \right] + j{\phi _{{m_x}{m_z}}}} \right\}} } .
\end{split}
\end{align}
\end{small}
\hrulefill
\end{figure*}
To maximize ${\left| \chi \right|^2}$, the optimal phase shift on the $\left( {{m_x},{m_z}} \right)$-th element can be written as
\begin{equation}
\begin{split}
& {\phi _{{m_x},{m_z}}} = k\left( {{m_z} - 1} \right)\left( {\cos {\varphi _{out,1}} - \cos {\varphi _{in,1}}} \right) \\
& + k\left( {{m_x} - 1} \right)\left( {\sin {\theta _{out,1}}\sin {\varphi _{out,1}} - \sin {\theta _{in,1}}\sin {\varphi _{in,1}}} \right).
\end{split}
\end{equation}
\par The above derivation is able to apply to every NLoS UE, the reflection matrix of RIS can be expressed as a function of angles as $\mathbf{\Phi}  = \mathbf{\Phi} \left(
{\theta ,\varphi } \right)$. By introducing this equation into (11), we can acquire
\begin{equation}
  \textbf{f} = \textbf{f}\left( {\theta ,\varphi ,\mathbf{\Phi} \left( {\theta ,\varphi } \right)} \right) = \textbf{f}\left( {\theta ,\varphi } \right),
\end{equation}
\begin{equation}
  \textbf{w} = \textbf{w}\left( {\theta ,\varphi ,\mathbf{\Phi} \left( {\theta ,\varphi } \right)} \right) = \textbf{w}\left( {\theta ,\varphi } \right).
\end{equation}
Further substituting (19), (20) and $\mathbf{\Phi}  = \mathbf{\Phi} \left( {\theta ,\varphi } \right)$ into (14), the optimization problem is then updated to
\begin{equation}
\begin{array}{ll}
& \mathop {\max }\limits_{{\bf{\Theta }},{\bf{\Psi }},{{\bf{\Theta }}_{RIS}},{{\bf{\Psi }}_{RIS}},{\bf{P}}} R, \\
& {\rm{s.t.}}\left\{ \begin{array}{l}
{\rm{C1}}:{{\vartheta _n} \in \left\{ {0,1} \right\},\forall n \in {\cal N},}\\
{\rm{C2}}:{0 \le {\theta _{R,l}},{\theta _{T,l}},{\theta _{in,l}},{\theta _{out,l}} \le 2\pi ,\forall l \in {\cal L} ,}\\
{\rm{C3}}:{0 \le {\varphi _{R,l}},{\varphi _{T,l}},{\varphi _{in,l}},{\varphi _{out,l}} \le 2\pi ,\forall l \in {\cal L} ,}\\
{\rm{C4}}:{{\bf{\Phi }} = f\left( {{{\bf{\Theta }}_{RIS}},{{\bf{\Psi }}_{RIS}}} \right),}\\
{\rm{C5}}:{0 < \sum\limits_{n = 1}^N {{P_n}}  \le {P^{\max }},}\\
{\rm{C6}}:{\left| {{{\bf{f}}_n}} \right|^2} = 1 ,\forall n \in {\cal N}\\
{\rm{C7}}:{\left| {{\textbf{w}_n}} \right|^2} = 1,\forall n \in {\cal N}.\\
\end{array} \right.
\end{array}
\end{equation}
where specific expression of the function $f\left(  \cdot  \right)$ in C4 is relevant to equation (18). It is not difficult to see that the transformed optimization problem no longer needs to optimize variable ${\bf{\Phi }}$ as the original constraint has been converted into an angle-dependent one, which can be directly calculated by ${{\bf{\Theta }}_{RIS}}$ and ${{\bf{\Psi }}_{RIS}}$.
\par It is worth mentioning that this work mainly investigates the beam alignment problem. Thus, for simplicity, we assume that the mBS allocates power to $N$ UEs evenly. The transmit power of UE $n$ is ${P_n} = {P_m} = \frac{{{P^{\max }}}}{N}$. Furthermore, denoting $p = {P_n},\forall n \in {\cal N}$ be the average transmit power of all UEs. By introducing $p$ into the formulated problem (21), it can be rewritten as
\begin{equation}
\begin{array}{ll}
& \mathop {\max }\limits_{{\bf{\Theta }},{\bf{\Psi }},{{\bf{\Theta }}_{RIS}},{{\bf{\Psi }}_{RIS}},{\bf{\Phi }}} \sum\limits_{n = 1}^N {\log_2 \left( {1 + \frac{{{{\left| {{\bf{w}}_n^H{{\bf{\Omega }}_n}{{\bf{f}}_n}} \right|}^2}{p}}}{{\sum\limits_{m \ne n}^N {{{\left| {{\bf{w}}_n^H{{\bf{\Omega }}_n}{{\bf{f}}_m}} \right|}^2}{p}}  + \sigma _w^2}}} \right)}, \\
& {\rm{s.t.}} \quad{\rm{C1}},{\rm{C2}},{\rm{C3}},{\rm{C4}},{\rm{C6}},{\rm{C7}}.
\end{array}
\end{equation}
\par In Fig. 1, NLoS UEs suffer from remarkable intra-cell interference since the mBS and the RIS are sole and fixed leading to a situation that several NLoS signals transmitting in nearly one direction between the mBS and RIS path. Therefore, to improve the system performance, we mean to suppress the interference through exploiting the zero-forcing beamforming (ZF-BF) at UE \cite{1261332}. The ZF-BF can force the interference mixed in receive signals to be zero and thus extract the needed signal alone.

\emph{Theorem}: Given ${\bf{W}} = {\left( {{\bf{w}}_1^H,...,{\bf{w}}_N^H} \right)^H} \in {\mathbb{C}^{N \times {N_R}}}$ denoted the ZF-BF matrix, where ${\mathbf{w}_n}$ is the ZF-BF vector at UE $n$. The ZF-BF matrix $\bf{W}$ can be obtained by the channel information and transmit beamforming vectors as
\begin{equation}
  {\bf{W}} = {\left( {{{{\bf{\hat F}}}^H}{{\bf{\Lambda }}^H}{\bf{\Lambda \hat F}}} \right)^{ - 1}}{{\bf{\hat F}}^H}{{\bf{\Lambda }}^H},
\end{equation}
where ${\bf{\hat F}} = {\left( {{\bf{F}},...,{\bf{F}}} \right)^H}$, ${\bf{F}} = \left( {{{\bf{f}}_1},...,{{\bf{f}}_N}} \right)$, ${\bf{\Lambda }} = \left( {{{\bf{\Lambda }}_1},...,{{\bf{\Lambda }}_N}} \right)$, ${{\bf{\Lambda }}_n} = \sqrt p \left[ {{\vartheta _n}{\bf{H}}_{n,LoS}^H + \left( {1 - {\vartheta _n}} \right){\bf{H}}_{n,NLoS}^H} \right]$.

\emph{Proof}: Taking UE $n$ as an example, the received signal at UE $n$ is described as (1), yet, in order to make follow-up operations more visualized, we transform this equation into
\begin{equation}
\begin{split}
  & {\textbf{y}_n} = \sqrt p {\bf{w}}_n^H\left[ {{\vartheta _n}{\bf{H}}_{n,LoS}^H + \left( {1 - {\vartheta _n}} \right){\bf{H}}_{n,NLoS}^H} \right]\sum\limits_{m = 1}^N {{{\bf{f}}_m}{s_m}}  \\
  & + {\bf{w}}_n^H\eta.
\end{split}
\end{equation}
Then, the sum of the product of the transmitted beamforming vector and the corresponding signal are decomposed into the form of vector multiplication, i,e, $\sum\limits_{m = 1}^N {{{\bf{f}}_m}{s_m}}  = \left( {{{\bf{f}}_1},...,{{\bf{f}}_N}} \right){\left( {{s_1},...,{s_N}} \right)^H} = {\bf{F}}{{\bf{s}}^H}$. For briefness, we ignore the noise momentarily in the following process as it will make no difference here. By substituting ${{\bf{\Lambda }}_n}$, ${\bf{F}}$ and ${\bf{s}}$, the desired interference-free equation of received signal at UE $n$ is expressed as follow
\begin{equation}
\begin{split}
{y_n} = {\bf{w}}_n^H{{\bf{\Lambda }}_n}{\bf{F}}{{\bf{s}}^H} = {s_n}.
\end{split}
\end{equation}
\par However, the ZF-BF vector is hard to be solved directly. Thus, there is a need to handle the question from the solvable ZF-BF matrix instead. In order to solve the ZF-BF matrix, the multi-UEs received signals formula is introduced:
\begin{equation}
  {\bf{Y}} = {\bf{W\Lambda \hat F}}{{\bf{s}}^H} = {{\bf{I}}_N}{{\bf{s}}^H} = {{\bf{s}}^H},
\end{equation}
where ${\bf{Y}} = ({y_1},...,{y_N})^H$ is the vector of received signals, ${\bf{\Lambda }}$ is the concatenated channel matrix of all UEs, and ${\bf{\hat F}}$ is the cascade of $N$ transmit beamforming matrices ${\bf{F}}$. By supposing ${\bf{W}}{{\bf{\Lambda }}}{\bf{\hat F}} = {{\bf{I}}_N}$, where ${{\bf{I}}_N}$ is the $N$-dimension identity matrix, the elimination of interference can be completed and the equation (24) can be acquired. Then, the ZF-BF vector of UE $n$ can be obtained as the $n$-th row of ${\bf{W}}$. It is worth mentioning that the validity of this assumption is subject to certain preconditions, i.e., the ZF-BF vector exists only when the communication system satisfies the constraint ${N_R} > N - 1$ according to the null-space theory \cite{1261332}. In addition, ${\bf{W}}$ is expressed by ${{\bf{\Lambda }}}$ and ${\bf{\hat F}}$, which are functions of angles, meaning that the equation (20) is still valid, i.e., the ZF-BF vector is available through the angle information.
$\blacksquare$
\par Due to the interference suppression, the intra-cell interference is considered negligible, and the final optimization problem can be written as
\begin{equation}
\begin{array}{ll}
& \mathop {\max }\limits_{\mathbf{\Theta} ,\mathbf{\Psi} ,{{\bf{\Theta }}_{RIS}},{{\bf{\Psi }}_{RIS}}} \sum\limits_{n = 1}^N {\log_2 \left( {1 + \frac{{{{\left| {{\bf{w}}_n^H{{\bf{\Omega }}_n}{{\bf{f}}_n}} \right|}^2}{p}}}{{\sigma _w^2}}} \right)},  \\
& {\rm{s.t.}} \quad{\rm{C1}},{\rm{C2}},{\rm{C3}},{\rm{C4}},{\rm{C6}},{\rm{C7}}.
\end{array}
\end{equation}
\par The main purpose of this Section is to focus the optimization goal on the angle information. Then how to obtain the angle information is the main focus of the next Section.
\section{Transformer-based Angle Prediction Scheme}
In this section, we propose a novel transformer-based angle prediction scheme, i.e., the AM, to obtain angle information.

The AM is a mapping function that reflects the correlation between the geographic location information of UE and the corresponding beam alignment angle information. Especially, we denote such function as ${f_{AM}}\left(  \cdot  \right)$ and the location of UE $n$ as $\left( {{x_n},{y_n}} \right)$. If UE $n$ is located at a LoS area, the angle information can be obtained by $\left( {{{\boldsymbol{\theta}}_n},{\boldsymbol{\varphi}_n}} \right) = {f_{AM}}\left( {{x_n},{y_n}} \right)$; if not, angles referring to RIS will be generated together as $\left( {{{\boldsymbol{\theta}}_n},{\boldsymbol{\varphi}_n},{{\boldsymbol{\theta }}_n^{RIS}},{{\boldsymbol{\varphi }}_n^{RIS}}} \right) = {f_{AM}}\left( {{x_n},{y_n}} \right)$. Further, the AM is capable of supporting the simultaneous processing of multiple location information, which can be represented as $\left( {{\bf{\Theta }},{\bf{\Psi }},{\bf{\Theta }}_{RIS},{\bf{\Psi }}_{RIS}} \right) = {f_{AM}}\left( {\left( {{x_1},{y_1}} \right),...,\left( {{x_N},{y_N}} \right)} \right)$. The AM contains one UE classifier and two transformers. The process of angle prediction through the AM is divided into two stages: (1) the locations of UEs ${\bf{V}} = \left( {\left( {{x_1},{y_1}} \right),...,\left( {{x_N},{y_N}} \right)} \right)$ are input into the UE classifier, which divides the UEs into two groups: LoS UEs and NLoS UEs, (2) the categorized UE locations are sent to the corresponding transformer, then the angle information for each UE is generated through transformers. It is worth noting that one transformer is also sufficient to forecast the angle information of UEs, however, the structure of two transformers performs higher accuracy. The UE classifier considered in this paper is a simple binary classifier, which is utilized for determining whether UE is blocked and to package the classification results into two UE sets. In the following subsections, we will provide a detailed introduction to the data preprocessing, attention mechanism, and encoding and decoding of our transformer model.
\begin{figure}[t!]
\begin{small}
	\centering
	\includegraphics[width=0.45\textwidth]{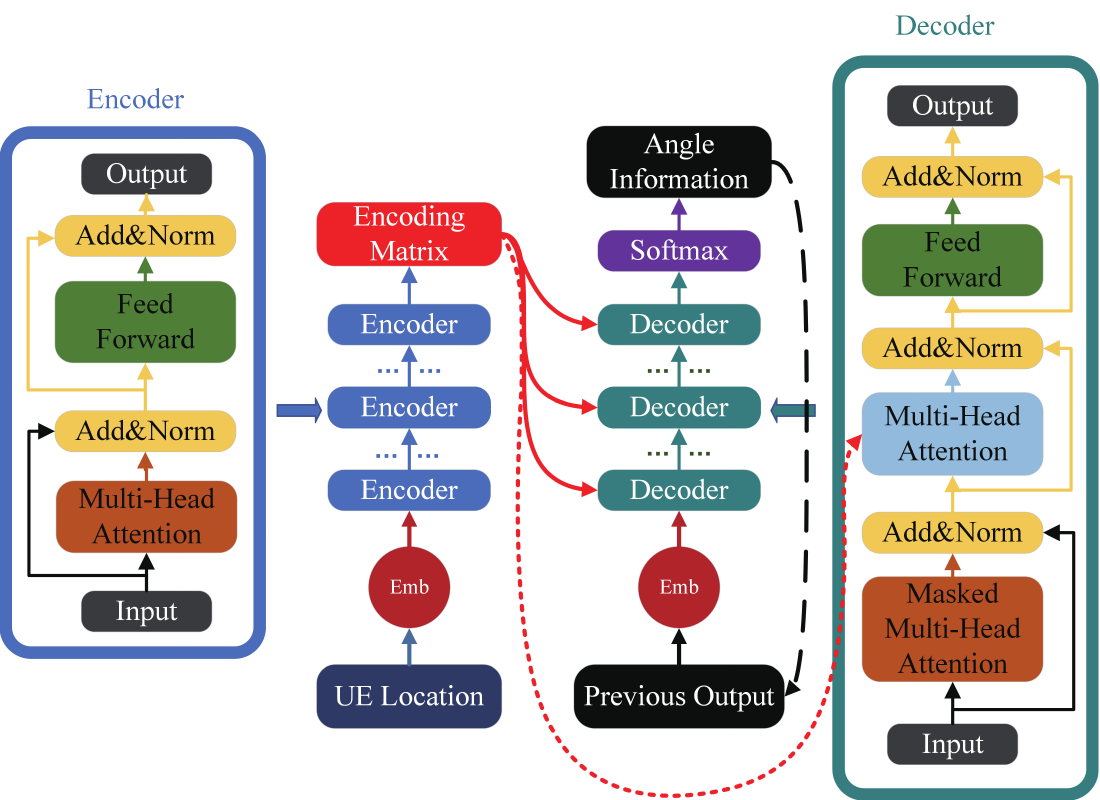}
	\caption{The structure of the transformer.}
	\label{fig:2}
\end{small}
\end{figure}
\subsection{Data Preprocessing}
The transformer algorithm is a deep learning model designed for disposing of sequence-to-sequence tasks (the UE locations-to-angles problem can also be regarded as one of such issue). The structure of the transformer model and the process of angle prediction are all shown in Fig. 2, it can be seen that the transformer model is composed of several encoders and decoders and the input of both the first decoder and encoder passes through an embedding layer. Such data preprocessing process in the transformer is called embedding. Take location information as an example, the embedding operation in the transformer model is mainly used to map discrete location information into a continuous vector space so that neural networks can process the location data more conveniently. Such process can be seen as mapping each value on the x-y-z axes of location data to a high-dimensional real-valued vector, and these vectors' features can represent the `semantic' relationships and `syntactic' structures between position coordinates. Finally, each input element of the model will be embedded as a vector with $d_e$ feature dimension and then fed into the encoder or decoder. As for the angle information, the same procedure will be executed until every angle be transformed into a embedding vector.
\subsection{Attention Mechanism}
The attention mechanism is the core concept of the transformer as shown in Fig. 2. In this part, we will introduce the attention mechanism from the self-attention to multi-head attention based on the general attention model. The input of an attention module is the output of the previous layer or the embedded UE location matrix, we denote the input matrix as ${{\bf{I}}_{att}} \in \mathbb{R}^{N \times {d_e}}$. Once the matrix ${{\bf{I}}_{att}}$ is input, it will be transformed into three new matrices at the first step in attention, which can be written as ${\bf{Q}} = {{{\bf{I}}_{att}{\bf{W}}_q}}$, ${\bf{K}} = {{{\bf{I}}_{att}{\bf{W}}_k}}$, and ${\bf{V}} = {{{\bf{I}}_{att}{\bf{W}}_v}}$, where ${d_k}$ is the feature dimension of the key matrix ${\bf{K}} \in {\mathbb{R}^{N \times {d_k}}}$ and query matrix ${\bf{Q}} \in {\mathbb{R}^{N \times {d_k}}}$, ${\bf{V}} \in {\mathbb{R}^{N \times {d_e}}}$ is the value matrix. ${{\bf{W}}_q} \in {\mathbb{R}^{{d_e} \times {d_k}}}$, ${{\bf{W}}_k} \in {\mathbb{R}^{{d_e} \times {d_k}}}$, and ${{\bf{W}}_v} \in {\mathbb{R}^{{d_e} \times {d_e}}}$ are corresponding linear transformation matrices of $\mathbf{Q}$, $\mathbf{K}$, and $\mathbf{V}$ respectively with trainable weights. Then $\mathbf{Q}$ does the dot product with the transpose of $\mathbf{K}$ since the query matrix $\mathbf{Q}$ represents the elements for which we want to calculate attention weights and the key matrix $\mathbf{K}$ represents the elements that determine how much attention each query should give to different positions, the dot product between $\mathbf{K}$ and $\mathbf{Q}$ can measure the similarity and relevance between them. Subsequently, the dot result is sent to a softmax layer to count the attention score, which can be written as ${\bf{A}} = softmax\left( {\frac{{{\bf{Q}}{{\bf{K}}^T}}}{{\sqrt {{d_k}} }}} \right)$, where the softmax function aims to represent the attention score in the shape of probability because such a form makes it easier for the weighted average. In the attention score calculating formula, $\sqrt {{d_k}} $ is utilized as a scaling factor applied to prevent the dot product from being too large or too small. At last, the output of the self-attention module is obtained as the result of the dot product between the attention score $\mathbf{A}$ and the value matrix $\mathbf{V}$, which is written as ${\bf{O}}^{self}_{att} = softmax\left( {\frac{{{\bf{Q}}{{\bf{K}}^T}}}{{\sqrt {{d_k}} }}} \right){\bf{V}} = {\bf{AV}}$.
\par Once the outcomes of each self-attention are obtained, the output of multi-head attention can be calculated through ${{\bf{O}}} = {{{\bf{O}}^{mul}_{att}{\bf{W}}_{mul}}}$, where $h$ is the number of self-attentions in a multi-head attention, ${{\bf{O}}^{mul}_{att}} = \left[ {{{ {\bf{O}}^{self}_{att,1} }},...,{{ {\bf{O}}^{self}_{att,h} }}} \right] \in {\mathbb{R}^{N \times {hd_e}}}$ is the concatenation matrix of $h$ output matrices of self-attentions, ${{\bf{W}}_{mul}} \in {\mathbb{R}^{{hd_e} \times {d_e}}}$ is a trainable linear transformation matrix utilized to make the output dimension of the multi-head attention consistent with the input.
\subsection{Encode and Decode}
The Encode and Decode are respectively responsible for feature extraction and representation learning of input data, and generating output sequences using the feature representation learned by the encoding module. For Encode, we can see from Fig. 2 that this module consists of a plurality of linearly connected encoders, and each encoder contains three components: multi-head attention, add \& norm layer, and forward feedback layer. Multi-head attention has been illustrated in the previous subsection. The add \& norm layer mainly plays a role in maintaining the stability of training in the transformer model. This module consists of one normalization layer and one residual connection layer, which can be expressed as $LN{\left( {{{\bf{I}}_{AN}}} \right)_{ij}} = \frac{{{I_{ij}} - {\mu _j}}}{{\sqrt {\sigma _j^2 + \varepsilon } }}$ and ${{\bf{O}}_{AN}} = {{\bf{I}}_{prev}} + LN\left( {{{\bf{I}}_{AN}}} \right)$, where ${{\bf{I}}_{AN}}$ denotes the input matrix of the add \& norm layer, ${{\bf{I}}_{prev}}$ denotes the input matrix of the previous layer, ${I_{ij}}$ is the element on the $i$-th row and the $j$-th column of ${{\bf{I}}_{AN}}$, ${\mu _j}$ and $\sigma _j^2$ are the expectation and variance of the $j$-th column of ${{\bf{I}}_{AN}}$, $\varepsilon $ is a small constant which is used to prevent $\sqrt {\sigma _j^2 + \varepsilon } $ being zero, $LN{\left( {{{\bf{I}}_{AN}}} \right)_{ij}}$ denotes the result of the normalization of ${I_{ij}}$, $LN\left( {{{\bf{I}}_{AN}}} \right)$ is the output matrix of the layer normalization, and ${{\bf{O}}_{AN}}$ presents the outcome of the add \& norm layer. The forward feedback layer is a simple double-layer fully connected layer, the only thing worth noting in this module is that the first layer requires a Relu activation function yet the second layer does not require one, outputs of the two layers are ${{\bf{O}}_{FC,1}} = \max \left( {0,{{\bf{I}}_{FC}}{{\bf{W}}_1} + {{\bf{b}}_1}} \right)$ and ${{\bf{O}}_{FC,2}} = {{\bf{O}}_{FC,1}}{{\bf{W}}_2} + {{\bf{b}}_2}$, where ${{\bf{I}}_{FC}}$ denotes the input matrix of fully connection module, ${{\bf{O}}_{FC,1}}$ and ${{\bf{O}}_{FC,2}}$ denote the outcomes, ${{\bf{W}}_1},{{\bf{W}}_2} \in {\mathbb{R}^{{d_e} \times {d_e}}}$ and ${{\bf{b}}_1},{{\bf{b}}_2} \in {1 \times \mathbb{R}^{{d_e}}}$ are weight matrices and bias of two fully connection layers respectively. So far, all internal structures in the encoder have been introduced. At last, the Encode part will output an encoding matrix $\mathbf{C}$ in which the relationship among UE locations can be unambiguously understood and interpreted by the model through the repetitive data feature extraction and the attention module training.

Similar to the Encode, the Decode is also a cascade structure, which is composed of several decoders. From Fig. 2, it is not hard to find that each decoder involves one additional masked multi-head attention with a connected add \& norm layer. Due to the characteristics of the transformer, previously predicted angles are needed as one of the input parameters. Therefore, once generated, the output of the Decode will be sent to the decoder as one part of the input until training or prediction termination. Moreover, in order to prevent decoders from obtaining future information in the training stage, data needs to be masked before they are sent to the Decode. The masking process is represented as $Mask\left( {{\bf{I}}{}_{mask}} \right) = {\bf{I}}{}_{mask} + {\bf{tril}}\left( { - \inf } \right)$, where ${\bf{I}}{}_{mask}$ denotes the input matrix of the masked multi-head attention, ${\bf{tril}}\left( { - \inf } \right)$ is a lower triangular matrix with all elements in the lower triangular being negative infinity. It is worth noting that the key matrix and value matrix in the multi-head attention are calculated with the encoding matrix ${\bf{C}}$, while the query matrix is obtained from the output of the previous layer, which are ${{\bf{Q}}_{dec}} = {{\bf{W}}_{dec,q}}{{\bf{I}}_{dec}}$, ${{\bf{K}}_{dec}} = {{\bf{W}}_{dec,k}}{\bf{C}}$ and ${{\bf{V}}_{dec}} = {{\bf{W}}_{dec,v}}{\bf{C}}$, where ${{\bf{Q}}_{dec}}$, ${{\bf{K}}_{dec}}$ and ${{\bf{V}}_{dec}}$ are query matrix, key matrix and value matrix of decoders respectively, and ${{\bf{W}}_{dec,q}}$, ${{\bf{W}}_{dec,k}}$ and ${{\bf{W}}_{dec,v}}$ are corresponding weight matrices of them. ${{\bf{I}}_{dec}}$ is the input matrix of each decoder. Using the encoding matrix ${\bf{C}}$ to calculate ${{\bf{K}}_{dec}}$ and ${{\bf{V}}_{dec}}$ in the decoder mainly aims to enable decoders to use global information when predicting angles, and since the encoding matrix ${\bf{C}}$ may be different from matrices in decoders in terms of dimension, such operation is also capable of enabling the matrices alignment.

In summary, we train the transformer model with the input dataset where the features are the geographic locations of UEs and the labels are the corresponding perfect beam alignment angles. The Encode extracts features between UE location information to generate the encoding matrix, which is then fed into the Decode along with previously predicted angles to predict the next angle. When training is finished, these two transformers with specific weights form the AM. We only need to input the location information of UEs into it to quickly obtain the spatial angles corresponding to their beam alignment.
\section{Simulation}
\begin{figure}[t!]
	\centering
	\includegraphics[width=0.45\textwidth]{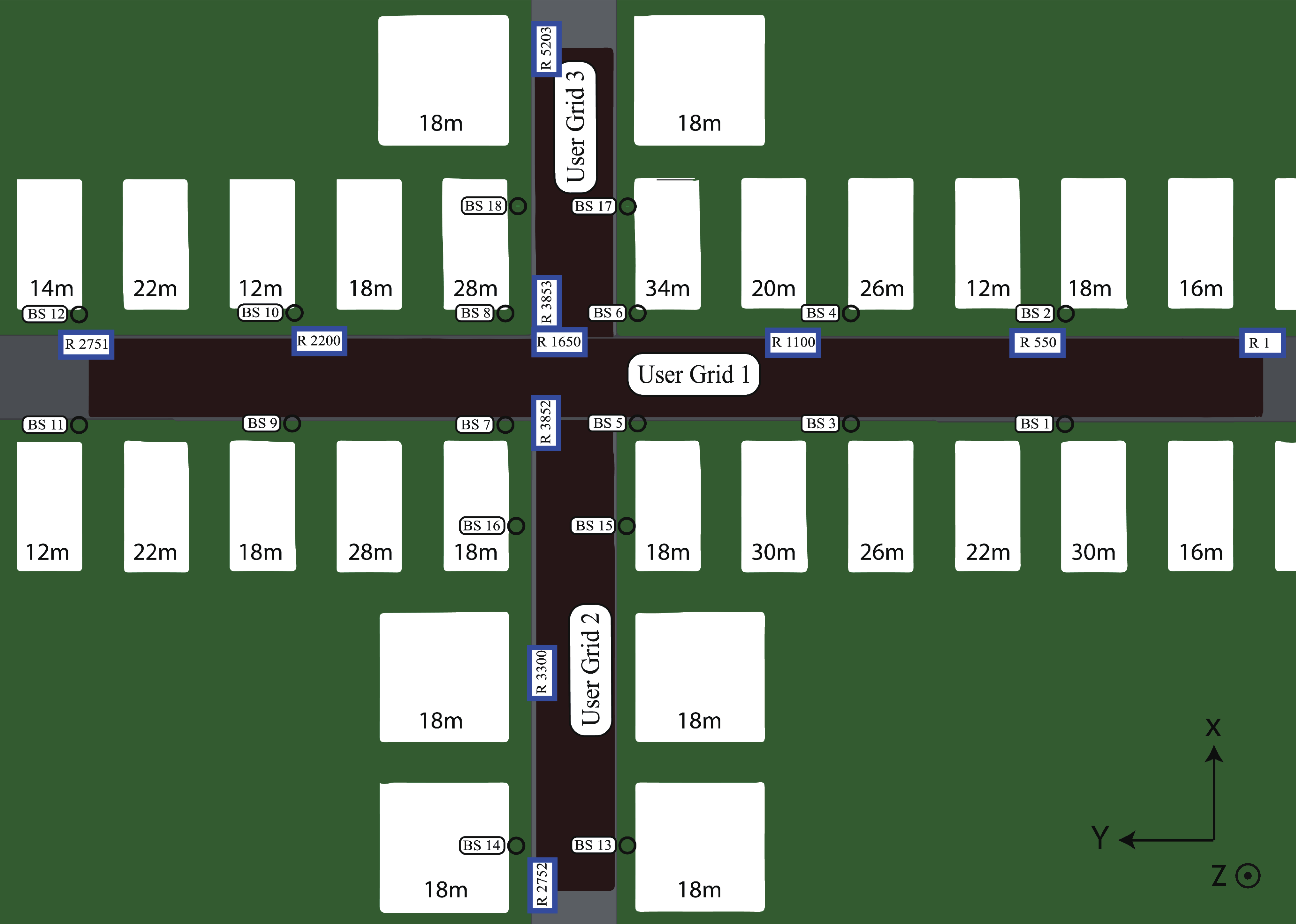}
	\caption{Scenario of O1\textunderscore 28.}
	\label{fig:3}
\end{figure}
\begin{figure}[t!]
	\centering
	\includegraphics[width=0.45\textwidth]{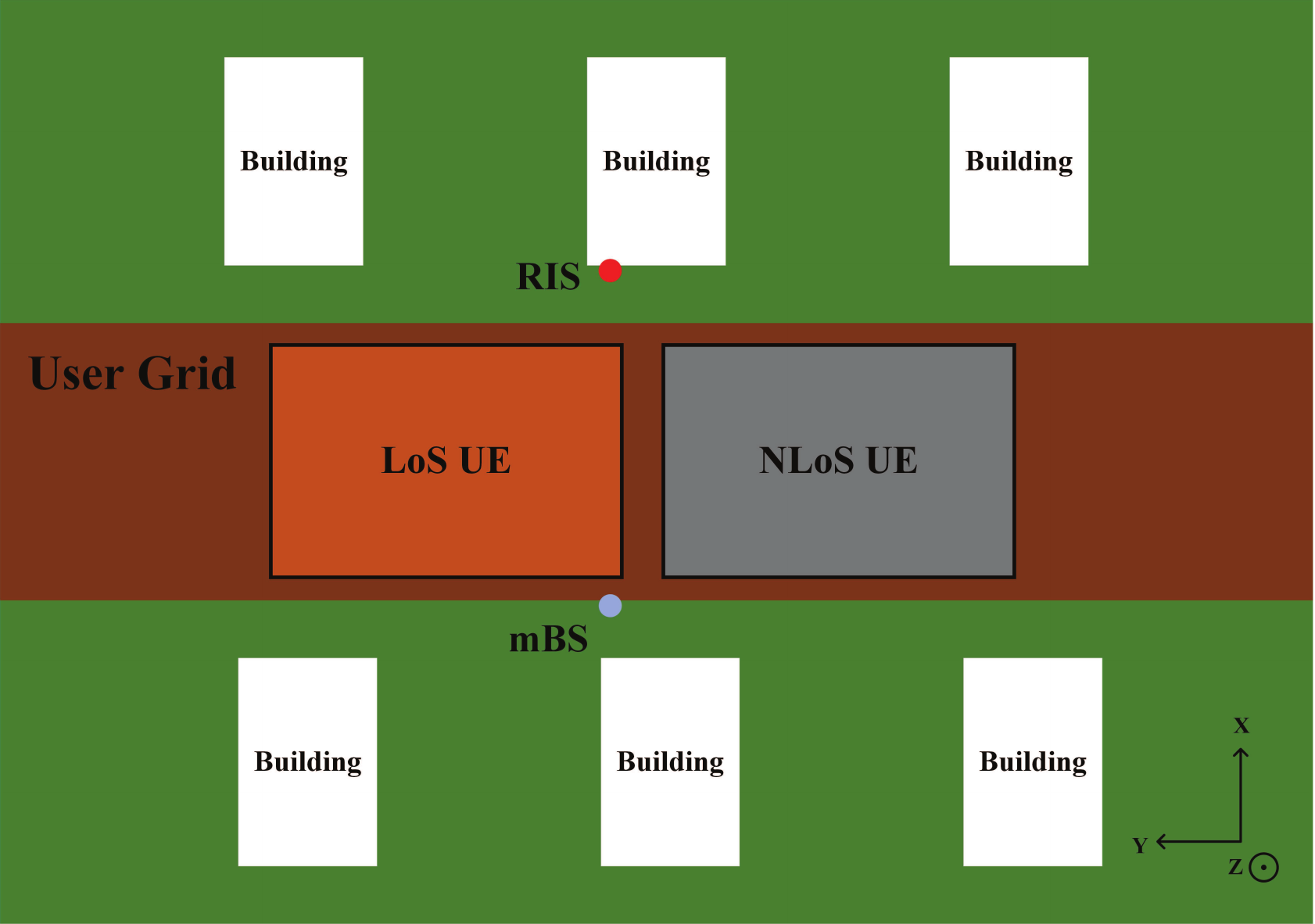}
	\caption{Diagram of the scenario in O1\textunderscore 28.}
	\label{fig:4}
\end{figure}
In this paper, simulations are conducted based on the widely used public dataset DeepMIMO \cite{alkhateeb2019deepmimo}, which is generated by a state-of-the-art commercial ray-tracing software called Wireless Insite \cite{remcom2019}. Specifically, we introduced the DeepMIMO `O1\textunderscore 28' scenario, which is shown in Fig. 3. We point out two UEs areas under the `O1\textunderscore 28', namely, LoS UEs zone and NLoS UEs zone. Moreover, two BSs in the scenario are activated for one transmits signals as mBS, and another assists communication with NLoS UEs as RIS \cite{9154301}. As shown in Fig. 4, BS 3 and BS 4 are respectively activated as mBS and RIS and fixed on the two sides of UE zones. Regarding the UE samples selection, we identify UEs from row \#$830$ to row \#$999$ as LoS UEs, and those from row \#$1000$ to row \#$1170$ are classified into NLoS UEs. All UEs and mBS are equipped with UPAs. Exactly, the mBS is equipped with an UPA of $4 \times 4$, each UE is equipped with an UPA of $2 \times 2$, and the RIS is considered as an UPA of $4 \times 4$. The antenna spacing of all devices is set as 0.5m. Moreover, signals are transmitted at the carrier frequency of 28 GHz, the total power of the mBS is 10dBm, and the noise power is -100dBm. We extract six datasets from the two pre-divided UEs regions (The number of both types of UEs in each dataset accounts for half), and the total sample number of these datasets are respectively 12240, 22236, 39440, 50100, and 61540. Before training the AM, datasets are further divided into training, validation, and test sets in a 3:1:1 ratio. Moreover, the Transformer model involved has 6 layers in both the Encoder and Decoder parts and 8 heads for each multi-head attention module. The training epoch is set as 100, and a stochastic gradient descent optimizer with 0.1 learning rate is considered.
\begin{figure}[t!]
	\centering
	\includegraphics[width=0.45\textwidth]{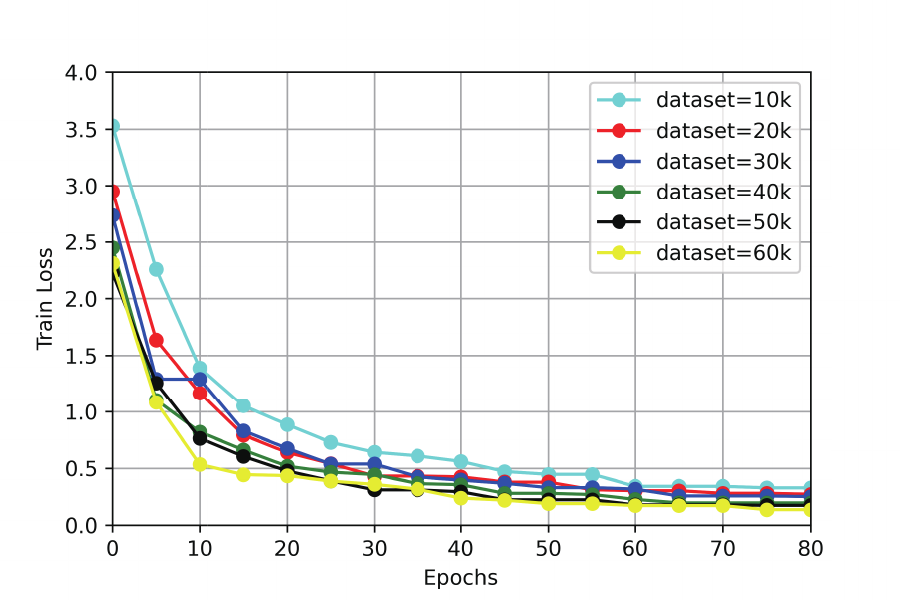}
	\caption{Convergency of proposed scheme.}
	\label{fig:5}
\end{figure}
\par From Fig. 5, it can be seen that the Transformer model is approximate to convergence around epoch 60 on all datasets. Informed by the train loss gap between epoch 20 and epoch 80 for each dataset, we can find that the larger the dataset, the faster the model converges. This is because large datasets usually have richer information diversity, which gives the model a broader understanding of the feature distribution, making it easier to find the correct optimization direction and reducing the number of iterations and oscillations. Moreover, it can be found from this figure that larger datasets result in lower final train losses, which generally implies better model performances. The main reason is that larger datasets typically contain more diverse samples and reflect a more comprehensive structure of the feature space, thereby reducing the probability of the model overfitting and enhancing the model's generalization ability. It is worth noting that the above-mentioned advantages of larger datasets come with the prerequisite that the additional data is of high quality and relevant to the task. If data quality is poor, noise is excessive, or there is significant distribution bias, larger datasets may actually slow down training and even degrade model performance.
\begin{figure}[t!]
	\centering
	\includegraphics[width=0.45\textwidth]{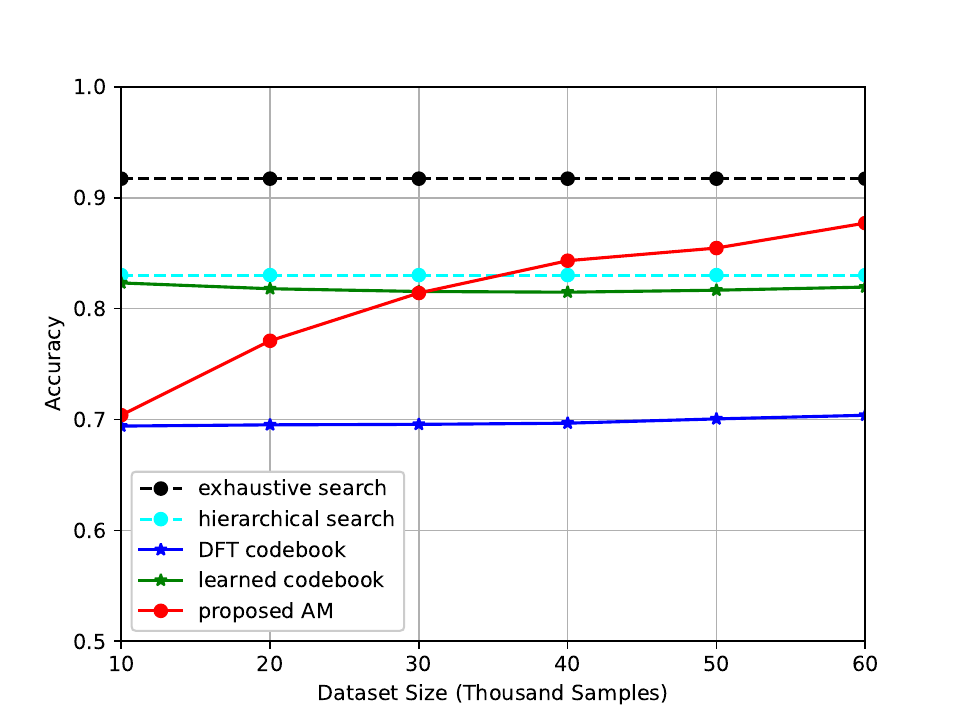}
	\caption{Beam alignment accuracy vs. dataset size.}
	\label{fig:6}
\end{figure}
\par Fig. 6 shows the average beam alignment accuracy of different beam alignment schemes. Except the proposed scheme, two search-based and two codebook-based beam alignment schemes are also examined as contrasts. To be specific, the two search-based methods refer to the exhaustive search and the two-layer hierarchical search, and the two codebook-based methods are the algorithms proposed in \cite{8999545} and \cite{9690703} respectively. It is worth mentioning that both codebook-based beam alignment algorithms utilize ML to improve the beam alignment efficiency. From Fig. 6, it can be observed intuitively that only the accuracy of the proposed scheme remarkably increases with the size of the dataset. By contrast, with the increasing dataset size, the accuracies of codebook-based schemes appear slightly different, and the accuracies of search-based schemes remain unchanged. Since the two search-based alignment methods are not driven by data, the accuracies of them are bound to be constants in this figure. The main reason for the significant difference among ML-based beam alignment algorithms is related to the types of models involved in the respective ML algorithms. The ML models mentioned in the two codebook-based schemes are convolutional neural network and multilayer perceptron, and the model involved in the proposed scheme is the Transformer. The Transformer usually has more parameters and stronger expressive power than the other two models, enabling it to capture more complex patterns in the data and resulting in better performance. However, the Transformer requires a large amount of data for training and may fall into overfitting if there is insufficient data. In contrast, simpler models, though limited in expressive power, tend to show more powerful generalization ability with fewer data. Therefore, when the two codebook-based schemes have already approached their expressive capacity limits on small datasets, the proposed scheme may not have fully learned the data features yet. This is why the accuracy of the proposed algorithm continues to increase as the dataset grows, while the accuracies of the codebook-based algorithms remain at a high level across all datasets. Only when the amount of data is large enough will the proposed scheme reach its performance limit. Thus, Fig. 6 shows that the proposed scheme's potential is gradually liberated as the dataset is enlarged. Finally, when the dataset reaches 60k, the beam alignment accuracy of the proposed scheme approaches that of the exhaustive search, which is a considerably high prediction level.
\begin{figure}[t!]
	\centering
	\includegraphics[width=0.45\textwidth]{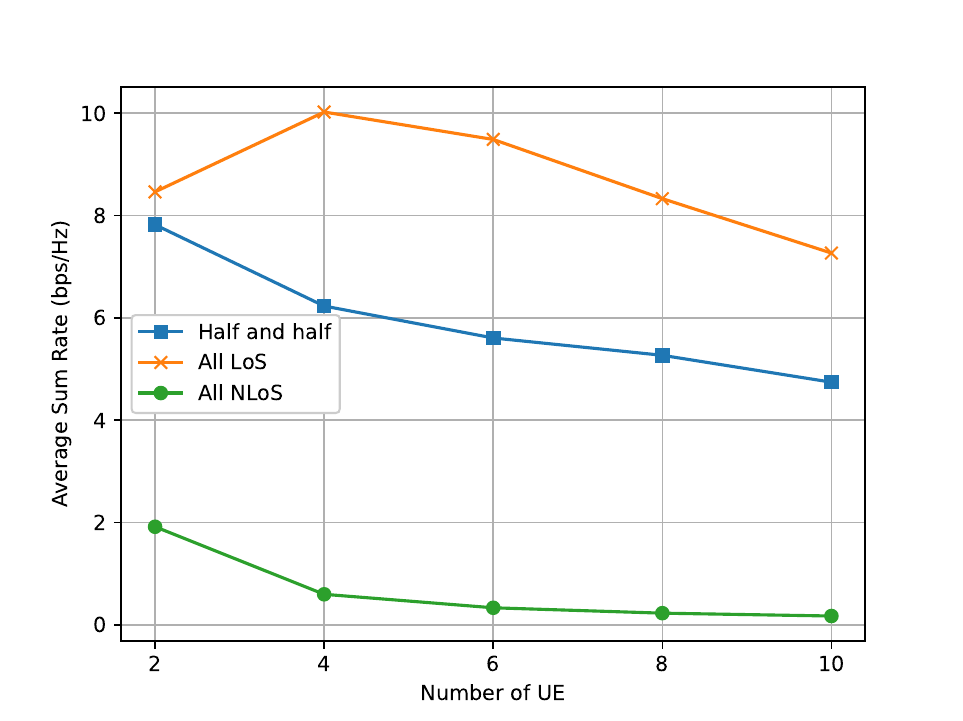}
	\caption{Average sum rate of proposed algorithm with different numbers of UE and different UE construction.}
	\label{fig:7}
\end{figure}
\par The average sum rate of the proposed algorithm under different UE numbers and different UE constructions is shown in Fig. 7. Three UE constructions are verified in this figure, i.e., all LoS UEs case, all NLoS UEs case, and LoS UEs and NLoS UEs each account for half. The dataset size is fixed at 60k. The horizontal axis represents the total number of UEs, and the vertical axis represents the sum of achievable rates. From the figure, it can be seen that regardless of the number of UEs, the sum rate of all NLoS is always the lowest, while that of all LoS is always the highest, which is due to the higher signal attenuation NLoS UE experienced. Moreover, as the total number of UEs increases, sum rates of all UE compositions show a declining trend. The main reason is that the allocated power for each UE keeps reducing as the number of UEs goes up while the interference is enhanced. It is necessary to mention that the intra-cell interference can not be ignored in this figure even if the ZF-BF is introduced to suppress it. This is because the null-space constraint may not be satisfied with the number of UEs increasing, which leads to the interference suppression performance reduction. Additionally, it can also be observed that the sum rate of the half and half case is generally higher than half the sum of the rates of all LoS case and all NLoS case. From Fig. 4, we know the LoS and NLoS areas are completely separated, and only one mBS and one RIS exist. That is to say, the interference from NLoS UEs to LoS UEs merely concentrated on one direction, i.e., the direction from the mBS to the RIS. Moreover, the mBS utilizes highly directional beams to transmit signals. Therefore, it is considered that most LoS UEs are almost impervious to the interference of NLoS UEs. Thus, the average mutual interference between LoS and NLoS UEs of the half and half case is the lowest among the three UE constructions.
\begin{figure}[t!]
	\centering
	\includegraphics[width=0.45\textwidth]{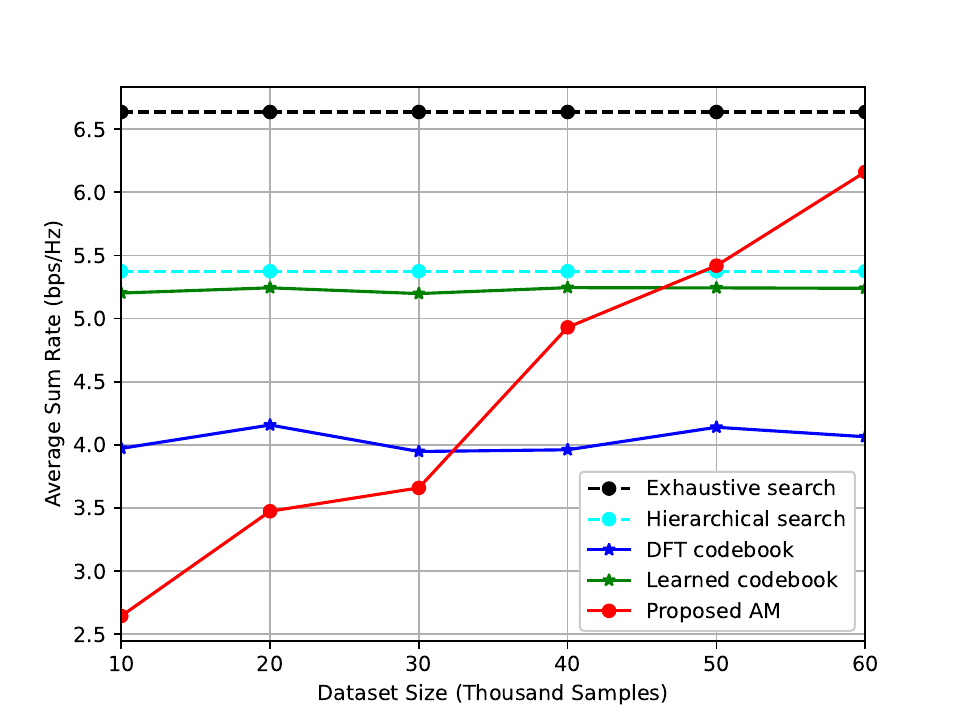}
	\caption{Average sum rate vs. dataset size.}
	\label{fig:8}
\end{figure}
\par Fig. 8 shows the average sum rate of different beam alignment algorithms with the size of the dataset. In this figure, the total number of UEs is fixed as 4 in each UE group, and 1000 groups of UEs are tested. Every UE group consists of half LoS UEs and half NLoS UEs. It can be seen from the figure that the distribution of the average sum rate of each algorithm is similar to the corresponding distribution of the accuracy in Fig. 4. That is, except for the proposed algorithm, the performance of other algorithms is almost unaffected by the size of the dataset. However, there are still noticeable differences, one of which is that the growth trend of the sum rate of the proposed algorithm becomes steeper with the increase in dataset size. In addition, when the dataset is small, even if the accuracy is almost the same as other algorithms, the sum rate of the proposed algorithm is significantly lower. These issues are caused by the inherent characteristics of the Transformer. As mentioned above, a large amount of data is essential to obtain a well-trained Transformer. Therefore the proposed scheme may generate larger prediction errors due to overfitting on small datasets. From the figure, although the codebook-based schemes show stabler performance on small datasets, the performance ceiling of the proposed scheme appears higher. The sum rate of the proposed algorithm rises rapidly as the dataset size increases and can eventually reach a level very close to the sum rate of the exhaustive search-based method.
\begin{figure}[t!]
	\centering
	\includegraphics[width=0.35\textwidth]{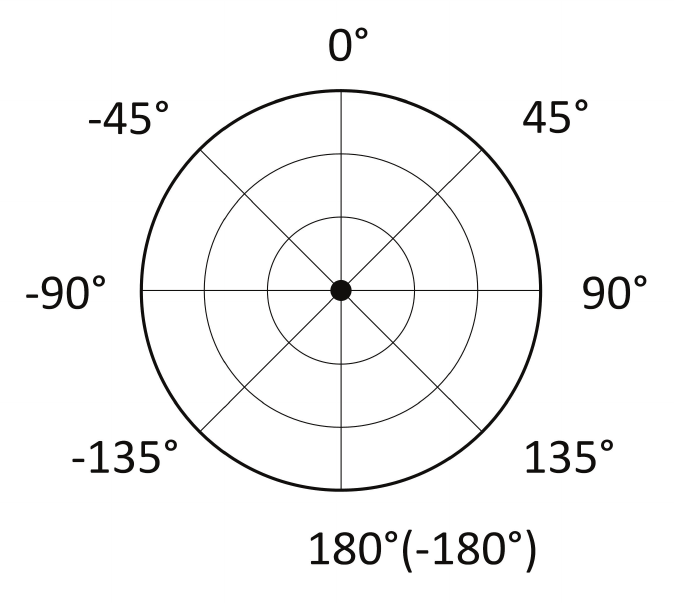}
	\caption{Polar coordinates of AM.}
	\label{fig:9}
\end{figure}
\begin{figure}[t!]
	\centering
	\includegraphics[width=0.45\textwidth]{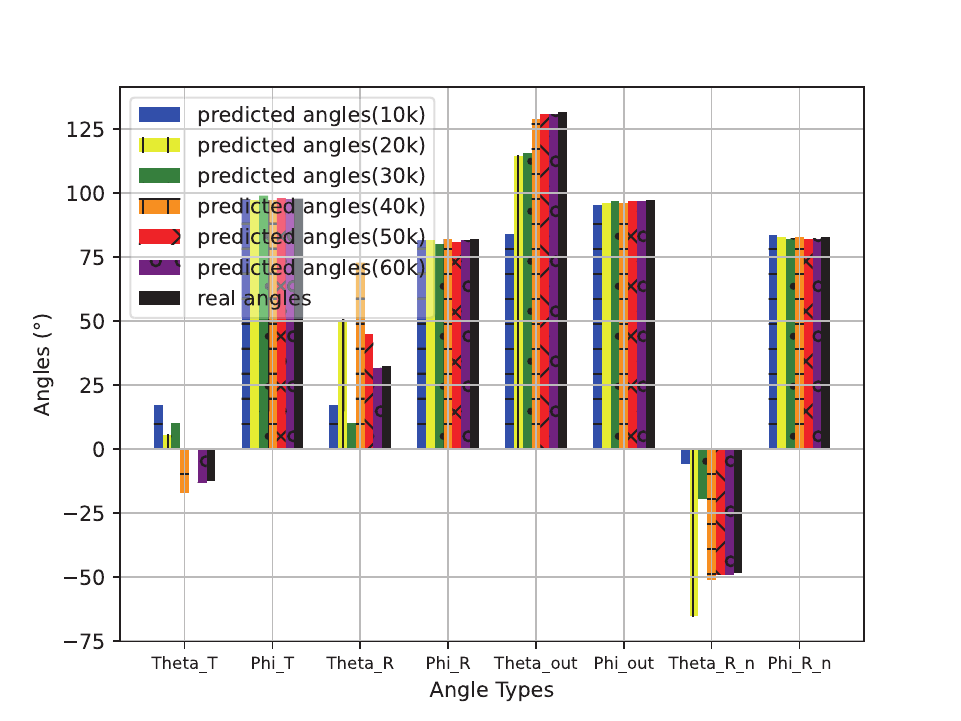}
	\caption{Real angles vs. predicted angles.}
	\label{fig:10}
\end{figure}
\par Fig. 9 shows a polar coordinate system with the mBS as the origin, which can provide reference coordinates for all angles involved in this work. Fig. 10 shows the comparison between the actual angles and predicted angles of the proposed scheme under different dataset sizes. Values of the predicted angles are calculated based on 1000 groups of UEs, with each group consisting of 2 LoS UEs and 2 NLoS UEs. The horizontal axis represents the type of angles, and the vertical axis represents the average angle value. To be specific, the first four angle types on the horizontal axis represent the horizontal AoD, the elevation AoD, the horizontal AoA, and the elevation AoA for LoS UEs, respectively. The latter four angle types represent the horizontal AoD, the elevation AoD, the horizontal AoA, and the elevation AoA from the RIS to NLoS UEs. It is worthy to mention that only the angles from the RIS to UEs are considered for NLoS UEs in this figure, as the beam alignment angles from the mBS to the RIS are constants. From Fig. 10, it can be seen that regardless of the dataset size, values of all elevation type predicted angles are very close to the corresponding actual angles. This is due to the value range of elevation angles is relatively small, around 30 degrees. Within such a small scale, plenty of duplicate and approximate angles will be recorded as data, whose features can be more easily learned by ML models. Thus, the Transformer can achieve high prediction accuracy in elevation angles on all datasets. In contrast, horizontal angles vary from 0 to 2$\pi$, which significantly contains more plentiful features. Thereby, the estimation accuracy of horizontal angles is more affected by the amount of data. Moreover, we can know from this figure that the larger the dataset, the smaller the gap between predicted and actual angles in all angle types in general. When dataset sizes are 10k, 20k, and 30k, the deviation between predicted and actual angles is relatively large, which explains why even though the proposed algorithm appears higher accuracy than the DFT codebook-based algorithm in Fig. 4, the average sum rate shows a worse performance in Fig. 8. The same is true for datasets of 40k and 50k, where the proposed algorithm clearly has higher accuracy than the learning codebook-based algorithm, but the average sum rate is lower. However, when the dataset reaches 60k, it can be seen that the predicted angles of the proposed algorithm closely match actual angles. Thus, in Fig. 8, the sum rate of the proposed algorithm is approximate to that of the exhaustive search algorithm.
\begin{figure*}[t!]
	\centering
    \subfigure[Real LoS AMs]{
        \includegraphics[width=0.4\textwidth]{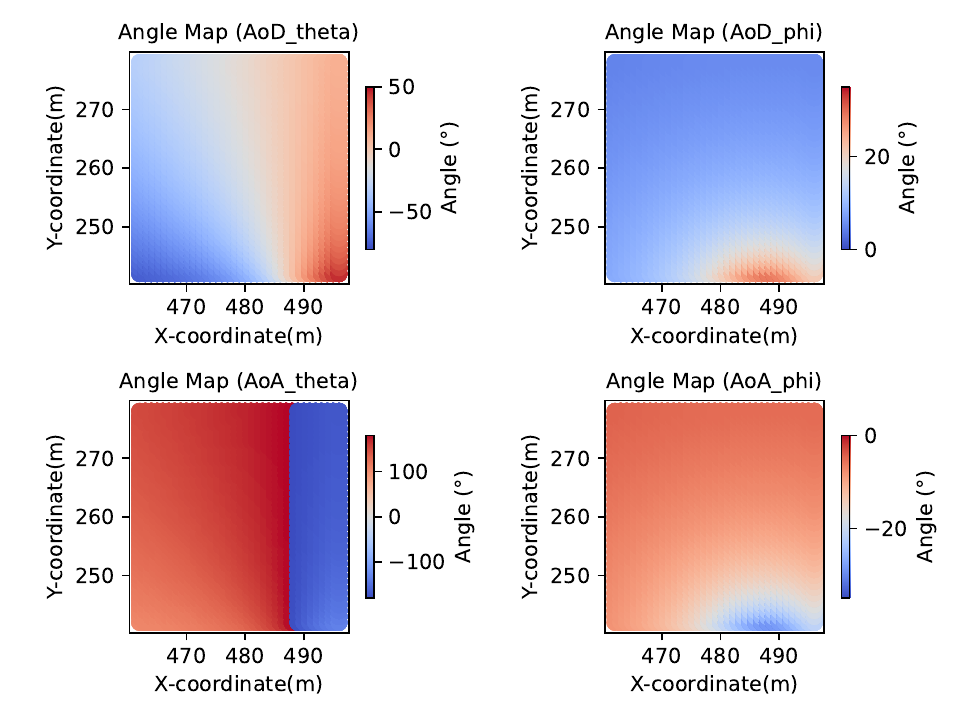}
    }
    \hspace{0.05\textwidth}
    \subfigure[Real NLoS AMs]{
        \includegraphics[width=0.4\textwidth]{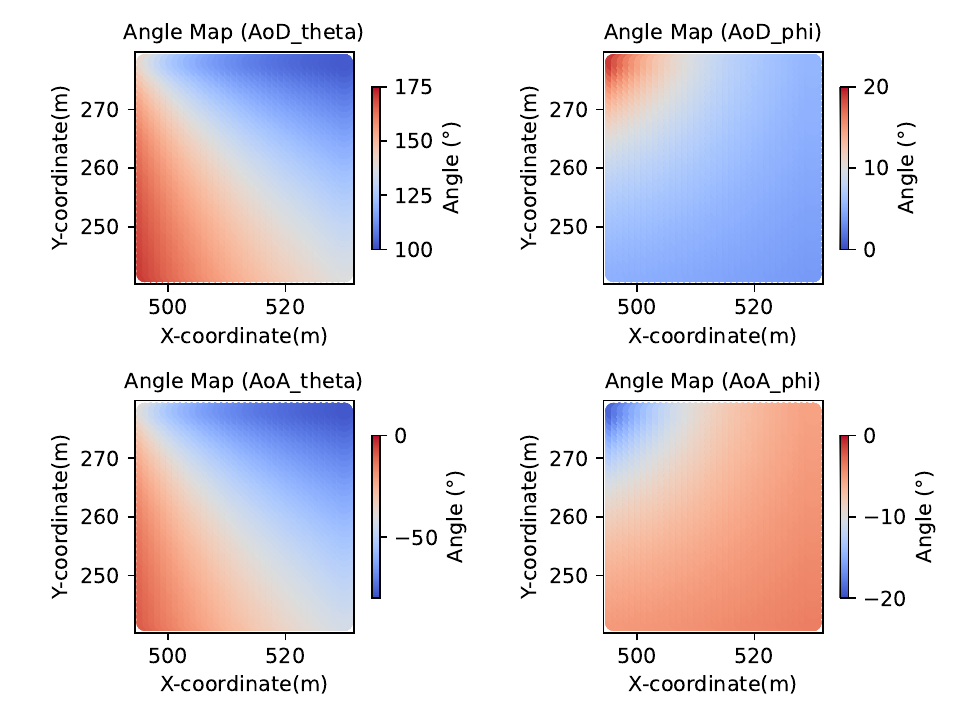}
    }
	\caption{Real AMs.}
	\label{fig:11}
\end{figure*}
\begin{figure*}[t!]
	\centering
    \subfigure[Predicted LoS AMs]{
        \includegraphics[width=0.4\textwidth]{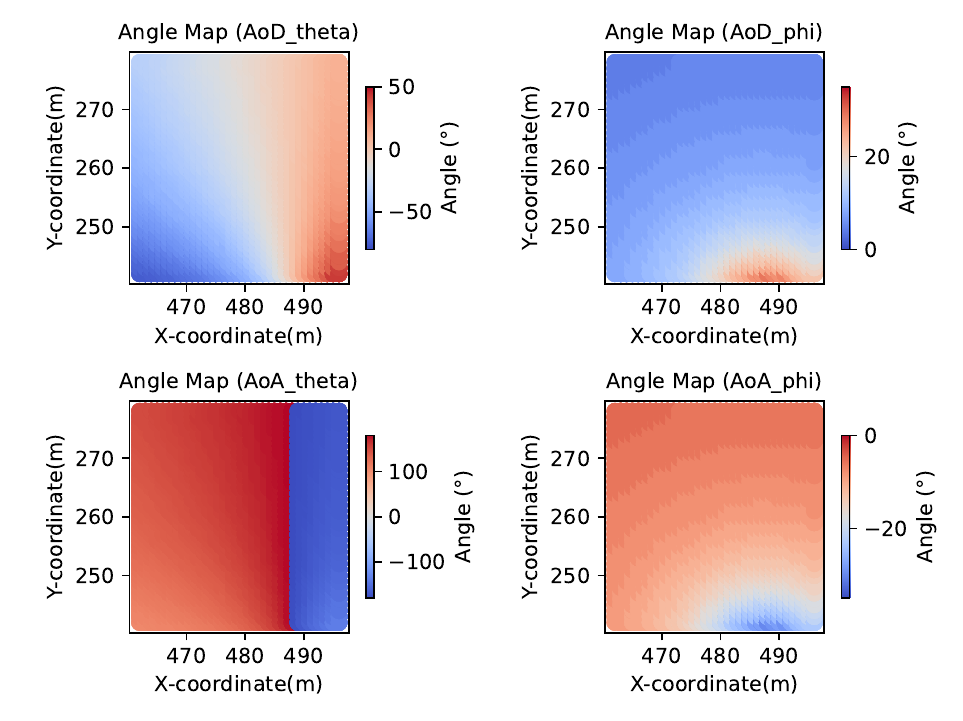}
    }
    \hspace{0.05\textwidth}
    \subfigure[Predicted NLoS AMs]{
        \includegraphics[width=0.4\textwidth]{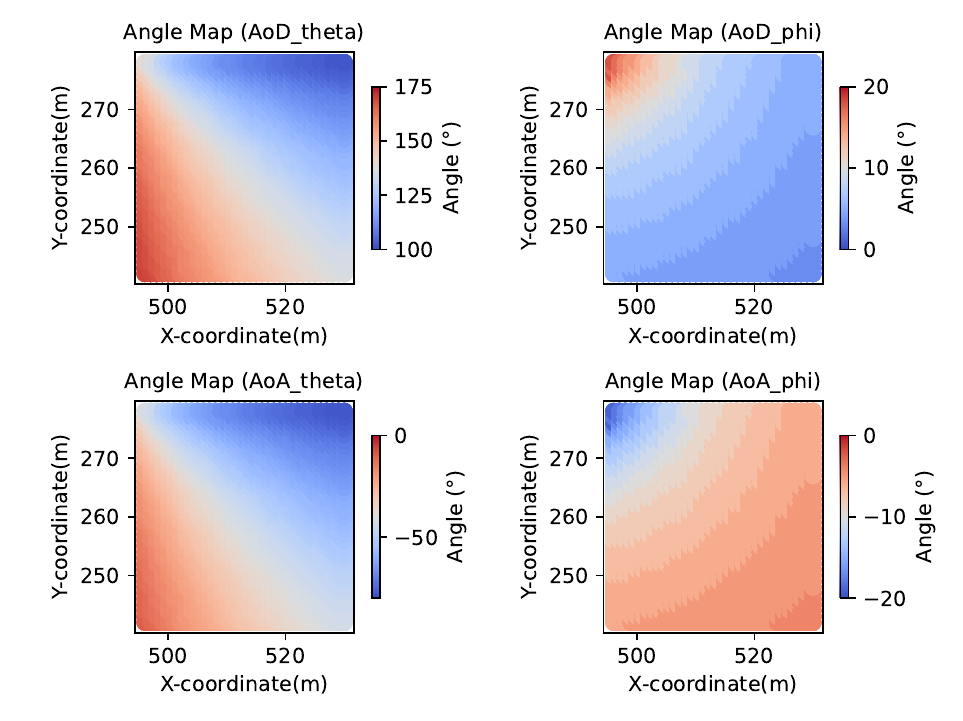}
    }
	\caption{Predicted AMs.}
	\label{fig:12}
\end{figure*}
\par Fig. 11 and Fig. 12 respectively show the actual AMs and predicted AMs when the dataset size is 60k. The horizontal and vertical coordinate of the AM represent the x-axis and y-axis in a rectangular coordinate system, which is defined in the DeepMIMO dataset. The coordinate points of the mBS and the RIS are (489.504, 235.504) and (489.504, 287.504), respectively. The subtitle of each map indicates the type of angles represented, and the color points on the map represent angle values. In AMs, each color point corresponds to the optimal beam alignment angle for the UE at that geographic location. It is worth mentioning that in NLoS AMs, all related angle types only refer to those between the RIS and NLoS UEs as the optimal beam direction between the mBS and the RIS is considered default. Through comparing Fig. 11 and 12, it is clear that predicted AMs are almost identical to actual AMs, with only minor differences in resolution. In addition, such distinction is more evident in elevation AMs, mainly due to the difference in data resolution. The precision of the practical data in the dataset is three decimal places. However, to reduce model parameters and improve training speed, we leverage approximated data to train the AM. Consequently, the sampling interval of the angle information used to train the AM is larger, and the precision of the angles predicted by the AM is bound to be slightly lower. This difference is more noticeable in a narrow angle range, which explains why the forecasted elevation AMs exhibit a reduced resolution while the predicted horizontal AMs show little difference. Regardless, from Fig. 10, we can see that even with data approximation, there is little impact on the prediction of beam alignment angles, and it has indeed significantly reduced computational overhead. From an intuitive perspective, the Transformer-based AM is capable of achieving impressive prediction accuracy.

\section{Conclusion}
In order to reduce the delay and the overhead caused by beam sweeping, we proposed a multi-UE beam alignment scheme, which utilizes the Transformer-based AM to predict optimal beam directions. This scheme can be effectively applied to RIS-aided mmWave communication systems through achieving fast and precise beam alignment for both LoS UEs and NLoS UEs. The performance and effectiveness has been well demonstrated in the simulation. Since implementing the proposed beam alignment scheme requires a vast amount of data, our next task may be investigating a way to alleviate the burden of data and considering the energy allocation issue jointly.

\ifCLASSOPTIONcaptionsoff
  \newpage
\fi

\bibliographystyle{IEEEtran}
\bibliography{reference}

\end{document}